\def\withFigures{1}  % if we want figures included, replace 0 -> 1
\documentclass[preprintnumbers,amsmath,amssymb, nofootinbib,tightenlines,preprint,12pt,superscriptaddress]{revtex4}
\usepackage{graphicx}

\usepackage{epsfig,amsmath,amssymb,verbatim,mathrsfs}
\newcommand{\beq}{\begin{eqnarray}}
\newcommand{\eeq}{\end{eqnarray}}
\newcommand{\bea}{\begin{eqnarray}}
\newcommand{\eea}{\end{eqnarray}}

\newcommand{\nnmb}{\nonumber}

\newcommand{\lrf}[2]{\left(\frac{#1}{#2}\right)}

\newcommand{\draftnote}[1]{\textbf{#1}}

\newcommand{\mev}{\text{MeV}}
\newcommand{\gev}{\text{GeV}}
\newcommand{\tev}{\text{TeV}}

\newcommand{\TRH}{T_\mathrm{RH}}
\newcommand{\Mpl}{M_\mathrm{Pl}}
\newcommand{\gzp}{{e^\prime}}
\newcommand{\blambda}{{\lambda^\prime}}
\newcommand{\bdlambda}{{\lambda^{\prime\prime}}}
\newcommand{\cp}{CP}
\newcommand{\cpt}{CPT}
\newcommand{\soft}{\mathrm{soft}}
\newcommand{\tr}{\mathrm{tr}}

\begin{document}

\setlength{\baselineskip}{0.2in}

\noindent

%\title{Dark Matter Antibaryons in a Supersymmetric Hidden Sector}
\title{Dark Matter Antibaryons from a Supersymmetric Hidden Sector}

\author{Nikita Blinov}
\affiliation{ Department of Physics and Astronomy, University of British Columbia,\\
Vancouver, BC V6T 1Z1, Canada}
\affiliation{Theory Group, TRIUMF, \\
4004 Wesbrook Mall, Vancouver, BC V6T 2A3, Canada}
\author{David E. Morrissey}
\affiliation{Theory Group, TRIUMF, \\
4004 Wesbrook Mall, Vancouver, BC V6T 2A3, Canada}
\author{Kris Sigurdson}
\affiliation{ Department of Physics and Astronomy, University of British Columbia,\\
Vancouver, BC V6T 1Z1, Canada}
\author{Sean Tulin}
\affiliation{Michigan Center for Theoretical Physics, Department of Physics,\\
University of Michigan, Ann Arbor, MI, USA, 48109}

\preprint{MCTP-12-13}

\begin{abstract}

The cosmological origin of both dark and baryonic matter can be explained through a unified mechanism called hylogenesis where baryon and antibaryon number are divided between the visible sector and a GeV-scale hidden sector, while the Universe remains net baryon symmetric.  The ``missing'' antibaryons, in the form of exotic hidden states, are the dark matter.  We study model-building, cosmological, and phenomenological aspects of this scenario within the framework of supersymmetry, which naturally stabilizes the light hidden sector and electroweak mass scales. Inelastic dark matter scattering on visible matter destroys nucleons, and nucleon decay searches offer a novel avenue for the direct detection of the hidden antibaryonic dark matter sea.

 \end{abstract}

\date{\today}

\maketitle

\setcounter{page}{2} %so that the .pdf file numbering matches the labels.

%%%%%%%%%%%%%%%%%%%%%%%%%%%%%%%%%%%%%%%%%%%%%%%%%%%%%%%%%%%%%%%%%%%%%%

\section{Introduction\label{sec:intro}}

The cosmological origin of both baryonic matter~\cite{Riotto:1999yt} 
and dark matter~\cite{Jungman:1995df} remain an important mystery in our
understanding of the early Universe.  An array of astrophysical observations 
indicates that a fraction $\Omega_{\rm b} \approx 4.6\%$ of the energy content
of the Universe is baryonic matter, while a fraction $\Omega_{\rm DM} \approx21\%$ 
is dark matter (DM)~\cite{Komatsu:2010fb}.
The Standard Model (SM) is incapable of explaining
these observations, providing no viable DM candidate, 
nor a successful mechanism for generating the baryon asymmetry.
Cosmology therefore requires new fundamental physics beyond the SM,
and it is important to find ways to detect such new physics experimentally.

The apparent coincidence between the densities of dark and baryonic matter, 
given by $\Omega_{\rm DM}/\Omega_{\rm b} \approx 5$, may be a clue that 
both originated through a unified mechanism.  A wide variety of models have been proposed along these lines within the framework of asymmetric DM~\cite{Nussinov:1985xr,Hooper:2004dc,Kitano:2004sv,Agashe:2004bm, Farrar:2005zd,Kohri:2009yn,Shelton:2010ta,Hut:1979xw,Dodelson:1989cq,Kuzmin:1996he,Gu:2007cw,An:2009vq, Davoudiasl:2010am}; see Ref.~\cite{Davoudiasl:2012uw} for a review.  In these scenarios, DM carries a conserved global charge, and its relic abundance is determined by its initial chemical potential.  Moreover, if the DM charge is related to baryon number ($B$), then the cosmic matter coincidence is naturally explained for $\mathcal{O}(5 \; {\rm GeV})$ DM mass.

In this work, we explore model-building, cosmological, and phenomenological aspects of hylogenesis (``matter-genesis''), a unified mechanism for generating dark and baryonic matter simultaneously~\cite{Davoudiasl:2010am,Davoudiasl:2011fj}.
Hylogenesis requires new hidden sector states that are neutral under
SM gauge interactions but carry non-zero $B$.  $\cp$-violating\footnote{$C$ is charge conjugation and $P$ is parity.} out-of-equilibrium decays in the early Universe generate a net $B$ asymmetry among the SM quarks and an equal-and-opposite
$B$ asymmetry among the new hidden states.  The Universe has zero total
baryon number, but for appropriate
interaction strengths and particle masses, the respective $B$ charges in 
the two sectors will never equilibrate, providing an explanation for
the observed asymmetry of (visible) baryons. The stable exotic 
particles carrying the compensating hidden antibaryon number produce the
correct abundance of dark matter.  Put another way, DM 
consists of the missing antibaryons.

\begin{figure}[ttt]
\begin{center}
\vspace{1cm}
\if\withFigures1
        \includegraphics[width = 0.18\textwidth]{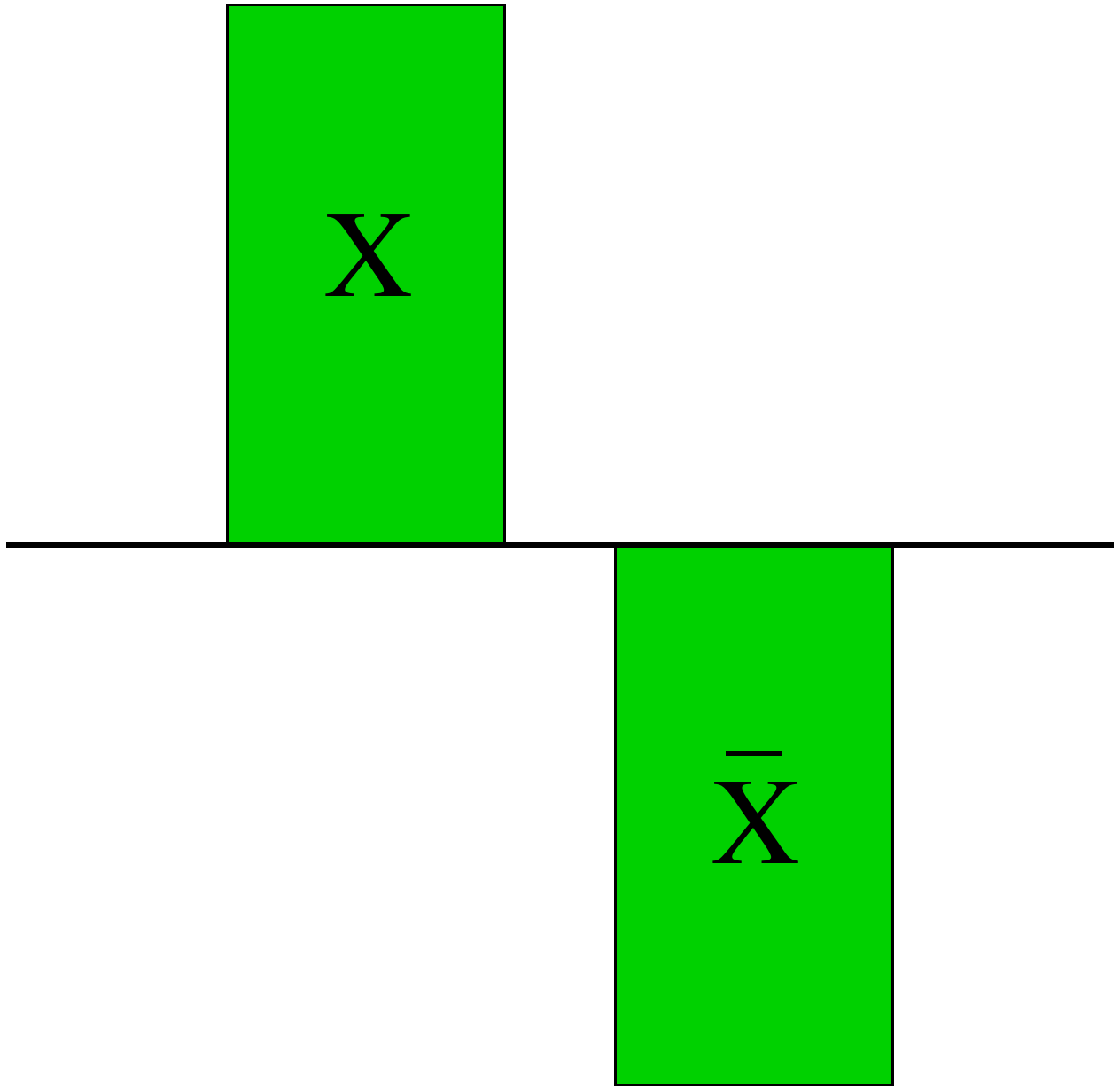}~~~~~
        \includegraphics[width = 0.36\textwidth]{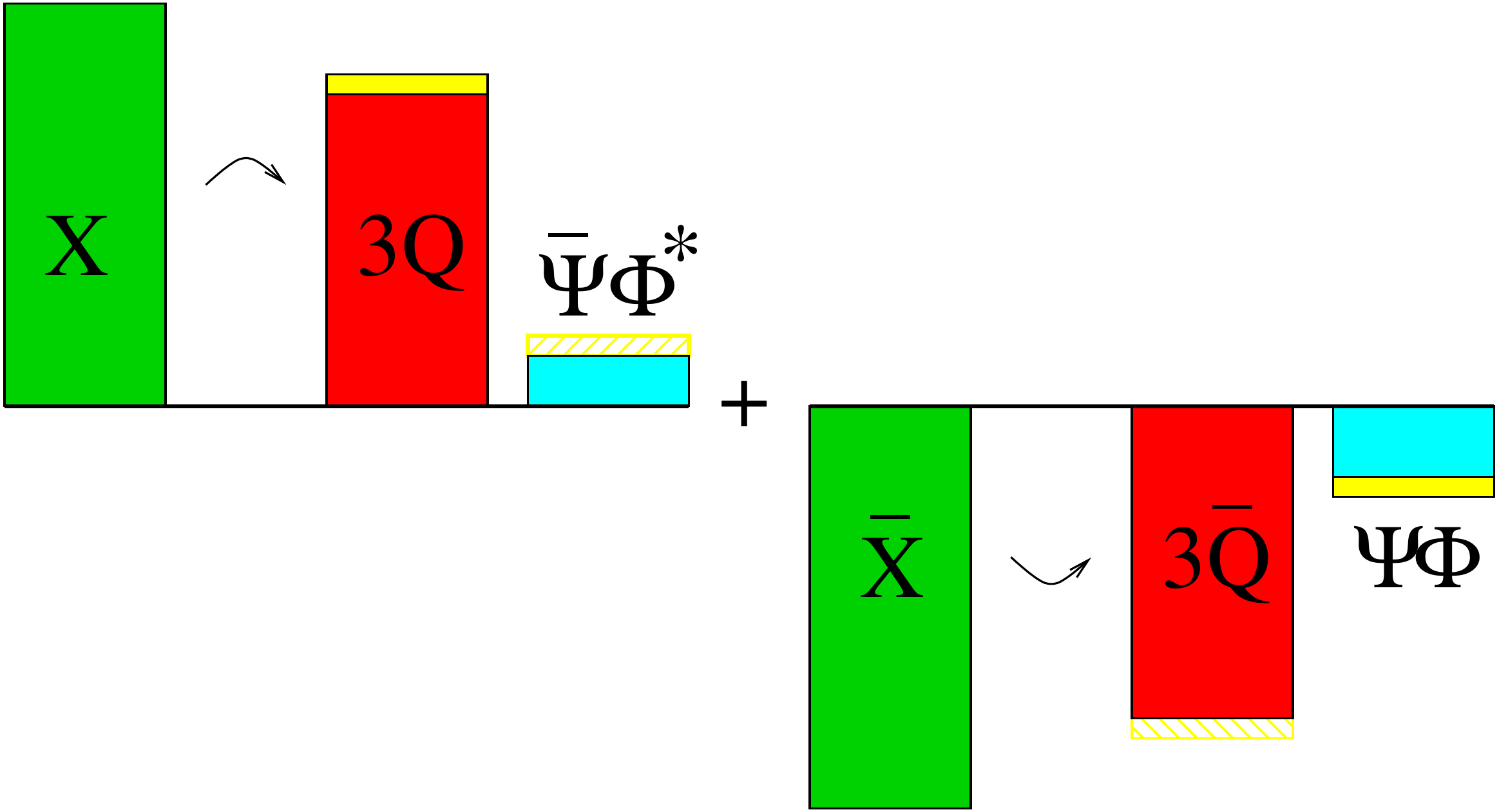}~~~~~
        \includegraphics[width = 0.36\textwidth]{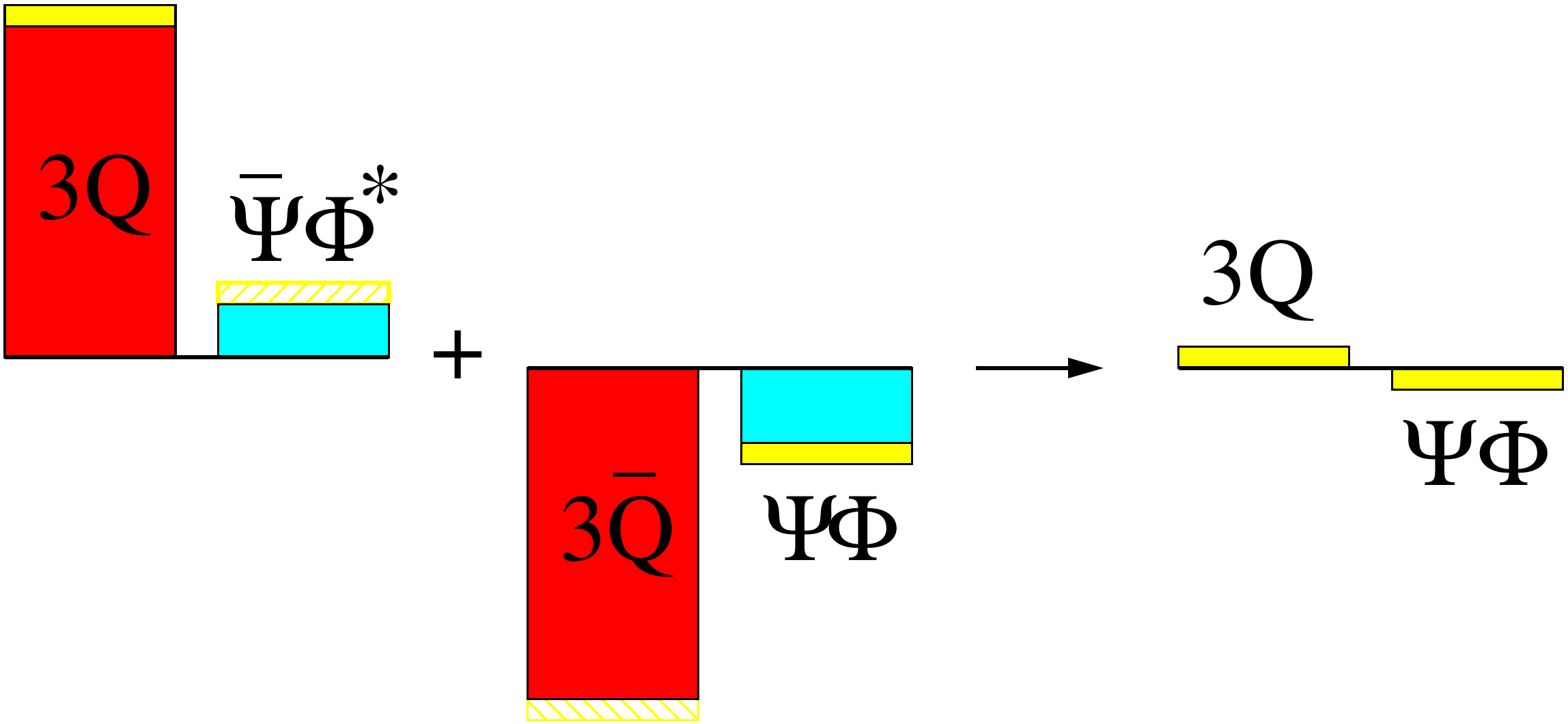}
\fi
\end{center}
\caption{(color online) The three steps of hylogenesis.}
\label{fig:hylopic}
\end{figure}

The minimal hylogenesis scenario, 
described in Refs.~\cite{Davoudiasl:2010am,Davoudiasl:2011fj},
has the following three ingredients: 
\begin{enumerate}
\item
DM consists of two states, a complex scalar $\Phi$ and Dirac fermion  $\Psi$, each carrying $B=-1/2$.
\item
A Dirac fermion $X$, carrying $B=1$, that transfers $B$ between quarks and DM through the gauge invariant operators~\cite{Dimopoulos:1987rk}
\beq
X\,u^c_{Ri}\,d^c_{Rj}\,d^c_{Rk} \, , \quad  X \Psi \Phi \label{eq:hylomin}
\eeq
where $i,j,k$ label generation (color indices and spinor contractions are suppressed).
\item An additional $U(1)^\prime$ gauge symmetry that is kinetically mixed with hypercharge and spontaneously broken near the GeV scale, producing a massive $Z^\prime$.  
\end{enumerate}
With these ingredients, hylogenesis proceeds in three stages,
which we illustrate schematically in Fig.~\ref{fig:hylopic}:
\begin{enumerate}
\item Equal ($\cp$-symmetric) densities of $X$ and $\bar{X}$
are created non-thermally, {\it e.g.}, at the end of a moduli-dominated
epoch when the Universe is reheated through moduli decay to a temperature
$\TRH$ in the range of 
$5\;\mev\lesssim \TRH \lesssim 100\;\gev \ll m_X$~\cite{Moroi:1999zb}.
\item The interactions of Eq.~\eqref{eq:hylomin} allow $X$ to decay to 
$u_{Ri}\,d_{R_j}\,d_{Rk}$
or $\bar{\Psi}\Phi^*$, and similarly for $\bar{X}$.  With at least
two flavours of $X$, these decays can violate $\cp$ leading to slightly
different partial widths for $X$ relative to $\bar{X}$,
and equal-and-opposite asymmetries for visible and hidden baryons.
\item Assuming $\Phi$ and $\Psi$ are charged under $U(1)^\prime$, 
the symmetric densities of hidden particles 
annihilate away almost completely, with $\Psi\bar{\Psi} \to Z'Z'$ 
and $\Phi\Phi^* \to Z'Z'$ occurring very efficiently in the hidden sector, 
followed by $Z'$ decaying to SM states via kinetic mixing.  
The residual antibaryonic asymmetry of $\Phi$ and $\Psi$ is asymmetric DM.
Likewise, the symmetric density of visible baryons and antibaryons annihilates 
efficiently into SM radiation.
\end{enumerate}
Both $\Psi$ and $\Phi$ are stable provided $|m_{\Psi}- m_{\Phi}| < m_p+m_e$, and they account for the observed DM density for $m_\Psi + m_\Phi \approx 5 m_p$, implying an allowed mass range $1.7 \lesssim m_{\Psi,\,\Phi} \lesssim 2.9$ GeV.  

On the phenomenological side, hylogenesis models possess a unique experimental signature: induced nucleon decay (IND), where antibaryonic DM particles scatter inelastically on visible baryons, destroying them and producing energetic mesons.  If $X$ couples through the ``neutron portal'' $u_R d_R d_R$, IND produces $\pi$ and $\eta$ final states, while if $X$ couples through the ``hyperon portal'' $u_R d_R s_R$, IND produces $K$ final states.  These signatures mimic regular nucleon decay, with effective nucleon lifetimes comparable to or shorter than existing limits; however, present nucleon decay constraints do not apply in general due to the different final state kinematics of IND.  Searching for IND in nucleon decay searches, such as the Super-Kamiokande experiment~\cite{Kobayashi:2005pe} and future experiments~\cite{Akiri:2011dv,IceCube:2011ac,Hewett:2012ns}, therefore offers a novel and unexplored means for discovering DM.

%%%%%%%%%%%%%%%%%%%%%%%%%%%%%%%%%%%%%%%%%%%%%%%%%%%%%%%%%%%%%%%%%%%%%%
%%%%%%%%%%%%%%%%%%%%%%%%%%%%%%%%%%%%%%%%%%%%%%%%%%%%%%%%%%%%%%%%%%%%%%

Although the minimal hylogenesis model described above successfully generates 
the cosmological baryon and DM densities, two puzzles remain.
Is there a natural framework to consider DM as a quasi-degenerate scalar/fermion pair?
Is there a mechanism to ensure the quantum stability of the GeV-scale
masses for hidden sector scalars?  Supersymmetry~(SUSY) 
can provide answers to both questions;
the DM pair $(\Phi,\Psi)$ forms a supermultiplet with $B=-1/2$, and
the stability of the GeV-scale hidden sector and 
the $(\Phi,\Psi)$ mass splitting is ensured naturally, provided 
SUSY breaking is suppressed in the hidden sector 
compared to the visible sector.

The goal of this work is to embed hylogenesis in a supersymmetric framework
of natural electroweak and hidden symmetry breaking, and to study in detail
the cosmological and phenomenological consequences.  
In Section~\ref{sec:susy}, we present a minimal supersymmetric extension 
of the hylogenesis theory described above.
We also address the origin of the nonrenormalizable nucleon portal operator 
%$X^c u_R^i d_R^j d_R^k$
$X\,u^c_{Ri}\,d^c_{Rj}\,d^c_{Rk}$. 
In Section~\ref{sec:baryo}, we investigate the cosmological dynamics 
of supersymmetric hylogenesis, showing explicitly the range of 
masses and parameters that can explain the correct matter densities.  
Section~\ref{sec:susybusting} contains a discussion
of how such parameter values can arise in a natural way from various
mechanisms for supersymmetry breaking.  In Section~\ref{sec:pheno} we
investigate the phenomenology of our model, including IND signatures, collider
probes, and DM direct detection.  
Our results are summarized in Section~\ref{sec:conc}. 
In Appendix~\ref{sec:hportal}, we also present an alternative supersymmetric
model based on Higgs portal mixing.

\section{Supersymmetric Hylogenesis Model
\label{sec:susy}}

In this section, we present an extension of the Minimal Supersymmetric
Standard Model~(MSSM) that can account for the dark matter and baryon
densities through a unified mechanism of hylogenesis.  In order to organize our
discussion, it is useful to divide our model into three sectors, given by the
superpotential terms
\beq
W = W_{\rm MSSM} + W_{\rm HS} + W_{\rm trans} \; .
\eeq
First, $W_{\rm MSSM}$ corresponds to the superpotential of the
usual MSSM with weak-scale superpartners; this is the visible sector.  
Second, we introduce a hidden sector comprised of new states which 
carry $B$, but are uncharged under the SM gauge group, 
and whose interactions are described by the superpotential $W_{\rm HS}$.  
The third term $W_{\rm trans}$ corresponds to operators responsible for 
$B$ transfer between the visible and hidden sectors.  
Baryon transfer operators generate equal-and-opposite $B$ asymmetries 
within the two sectors, and lead to IND signatures in nucleon decay searches.

\subsection{Hidden sector}

The hidden sector of our model consists of (i) four vector-like chiral superfields carrying nonzero $B$, 
denoted $X_{1,2}$ and $Y_{1,2}$, with charge-conjugate partners $X_{1,2}^c$ and $Y_{1,2}^c$,\footnote{Two species $X_{1,2}$ are required for $\cp$-violating decays (see Sec.~\ref{sec:baryo}), while two species $Y_{1,2}$ are needed
to couple them to the gauge-singlet $X$ fields.
%required in order to have a nonanomalous $U(1)^\prime$ gauge symmetry.
} and (ii) a $U(1)^\prime$ gauge sector, with gauge boson $Z^\prime$ and gauge coupling $\gzp$, spontaneously broken by a vector pair of hidden Higgs supermultiplets $H,H^c$.  Table~\ref{fields} summarizes these exotic fields.  The superpotential is given by
\begin{align}
W_{\rm HS} = \sum_{a=1,2} &\zeta_{a} X_{a} Y_1^2 + \bar{\zeta}_{a} X_{a}^c (Y^{c}_1)^2 + \gamma Y_1 Y_2^c H + \bar \gamma Y_1^c Y_2 H^c \notag \\
& \quad + {\mu}_{X_{a}} X_{a} X_{a}^c +  {\mu}_{Y_{a}} Y_{a} Y_{a}^c + \mu_H H H^c  \; , \label{WHS}
\end{align}
which includes Yukawa-type interactions with couplings $\zeta_{1,2}$, $\bar\zeta_{1,2}$, $\gamma$, $\bar \gamma$, and vector masses $\mu_{X_{1,2}}$, $\mu_{Y_{1,2}}$, $\mu_H$. We also assume a canonical K\"{a}hler potential for these multiplets.  Note as well that we have extended $R$-parity to a $\mathbb{Z}_4^R$ for the $Y_i^{(c)}$ multiplets.  Aside from allowing the couplings listed above, this extension does not lead to any novel features in the present case, beyond those imposed by the standard $R$-parity.

%%%%%%%%%%%%%%%%%%%%%%%%%%%%%%%%%%%%%%%%%%%5
\begin{table}[ttt]%[b!]
\begin{center}
\begin{tabular}{|cc|ccc|}
\hline
superfield && $U(1)^\prime$ & $B$ & $\; R\; $ \\
\hline
hidden baryons & $X_{1,2}$ & $0$ & $+ 1$ &  $-1$ \\
& $Y_1$ & $0$ & $- 1/2$ & $i$\\
& $Y_2$ & $+1$ & $- 1/2$ & $i$ \\
\hline 
hidden $U(1)^\prime$ & $H$ & $+1$ & $0$&  $+1$ \\
& $Z^\prime$ & $0$ & $0$& $0$ \\
\hline
\end{tabular}
\caption{\it New superfields in the hidden sector, with quantum numbers under $U(1)^\prime$, $B$, and $R$-parity.  
Chiral supermultiplets $X_{1,2}, Y_{1,2}, H$ also include vector partners
$X_{1,2}^c, Y_{1,2}^c, H^c$ with opposite charge assignments (not listed). \label{fields}}
\end{center}
\end{table}
%%%%%%%%%%%%%%%%%%%%%%%%%%%%%%%%%%%%%%%%%%%

  After symmetry breaking in the hidden sector, the superfields $Y_{1,2}$ and $Y_{1,2}^c$ mix to form two Dirac fermions $\Psi_{a}$ ($a =1,2$) and four complex scalars $\Phi_{b}$ ($b=1,2,3,4$) with $B=-1/2$.  Among these, the lightest states $\Psi_1$ and $\Phi_1$ are stable DM.  The fermionic mass terms for $Y_{1,2}$ are (in two-component notation)
\beq
\mathscr{L}_{\rm ferm} = - \big(Y^c_1, Y_2^c \big) \,\mathbf{M}_Y \left( \begin{array}{c} Y_1 \\ Y_2 \end{array} \right) + \textrm{h.c.} \; , \qquad
\mathbf{M}_Y \equiv \left( \begin{array}{cc} \mu_{Y_1} & \bar \gamma \eta_c \\ \gamma \eta  & \mu_{Y_2} \end{array} \right) \; .
\eeq
where $\eta \equiv \langle H \rangle$, $\eta_c \equiv \langle H^c \rangle$ are the hidden Higgs vacuum expectation values (vevs).  This mass matrix can be diagonalized by a biunitary transformation $V^T \mathbf{M}_Y U^\dagger = \mathrm{diag}(m_{\Psi_1},m_{\Psi_2})$.
The scalar mass terms in the basis $\widetilde Y \equiv ( \widetilde Y_1 ,\widetilde Y_2 , \widetilde Y_1^{c*} , \widetilde Y_2^{c*} )^T$ are
\beq
\mathscr{L}_{\rm scalar} = - \widetilde Y^\dagger \mathbf{M}_{\widetilde Y}^2 \widetilde Y \; .
\eeq
The $4 \times 4$ mass matrix $\mathbf{M}_{\widetilde Y}^2$ receives contributions from $F$ terms from Eq.~\eqref{WHS}, $D$ terms, and soft SUSY-breaking terms
\begin{align}
- \mathscr{L}_{\rm soft} \supset & \; m^2_{ Y_1} |\widetilde Y_1|^2 + m^2_{ Y_2} |\widetilde Y_2|^2  + m^2_{ Y^c_1} |\widetilde Y_1^c|^2 + m^2_{ Y^c_2} |\widetilde Y_2^c|^2 \\
& \quad + \left( b_1 \widetilde Y_1 \widetilde Y_1^c + b_2 \widetilde Y_2 \widetilde Y_2^c + \gamma A_\gamma \widetilde Y_1 \widetilde Y_2^c H + \bar{\gamma} A_{\bar \gamma} \widetilde Y_1^c \widetilde Y_2 H^c + \textrm{h.c.} \right)  \; , 
\end{align}
We have
\beq
\mathbf{M}^2_{\widetilde Y} \equiv \left( \begin{array}{cc} \mathbf{M}_Y^\dagger \mathbf{M}_Y - \boldsymbol{\delta} + \mathbf{m}_{\widetilde Y}^2 & \boldsymbol{\Delta}^\dagger \\ \boldsymbol{\Delta} & \mathbf{M}_Y \mathbf{M}_Y^\dagger + \boldsymbol{\delta} + \mathbf{m}^2_{\widetilde Y^c} \end{array} \right) \, , \quad 
\boldsymbol{\Delta} \equiv \left( \begin{array}{cc} b_1  & \gamma A_\gamma\eta \\  {\bar\gamma}A_{\bar \gamma} \eta_c & b_2 \end{array} \right) \; ,
\eeq
also defining $\mathbf{m}^2_{\widetilde Y} \equiv \textrm{diag}( m^2_{ Y_1}, m^2_{ Y_2} )$, $\mathbf{m}^2_{\widetilde Y^c} \equiv \textrm{diag}( m^2_{ Y_1^c}, m^2_{ Y_2^c} )$, $\boldsymbol{\delta} \equiv \gzp^2 (\eta_c^2 - \eta^2) \times \textrm{diag}(0,1)$.  The scalar mass matrix can be diagonalized by a unitary transformation $Z \mathbf{M}_{\widetilde Y}^2 Z^\dagger = \mathrm{diag}(m_{\Phi_1}^2,m_{\Phi_2}^2,m_{\Phi_3}^2,m_{\Phi_4}^2)$.  Similar mass matrices arise for the $X$ supermultiplets; for simplicity we assume that the fermion states $X_{1,2}$ and scalar states $\widetilde X_{1,2}$, $\widetilde X_{1,2}^c$ are all mass eigenstates.

%%%%%%%%%%%%%%%%%%%%%%%%
\begin{comment}
The mass eigenstates of the states $X_a$  are 
constructed in a similar manner. The fermion masses in two-component notation
are given by
\beq
\mathscr{L}_{\rm ferm} \supset -\big(X_1^c, X_2^c\big)\mathbf{M}_X 
\left( \begin{array}{c} X_1 \\ X_2 \end{array} \right) + \textrm{h.c.} \; , \qquad
\mathbf{M}_X = \mathrm{diag}(\mu_{X_1},\mu_{X_2})
\eeq
For the scalar components of $X_a$ multiplets, the relevant soft terms are
\beq
- \mathscr{L}_{\rm soft} \supset (\mathbf{m}^2_{\widetilde X})_{ab}\,\widetilde{X}_a^*\widetilde{X}_b
+ (\mathbf{m}^2_{\widetilde{X}^c})_{ab}\widetilde{X}_a^{c*}\widetilde{X}^c_b 
+ \left[(\mathbf{b}_{\widetilde{X}})_{ab} \widetilde{X}_a \widetilde{X}_b + \textrm{h.c.}\right]\, ,
\eeq
which give a $4\times4$ mass matrix 
\beq
\mathbf{M}^2_{\widetilde{X}} = \left(
\begin{array}{cc}
\mathbf{M}_X^\dagger \mathbf{M}_X + \mathbf{m}^2_{\widetilde{X}} & \mathbf{b}_{\widetilde{X}}^\dagger \\
  \mathbf{b}_{\widetilde{X}} & \mathbf{M}_X^\dagger \mathbf{M}_X + \mathbf{m}^2_{\widetilde{X}^c}
\end{array}
\right)
\eeq
in the $(\widetilde{X}_1,\widetilde{X}_2,\widetilde{X}_1^{c*},\widetilde{X}_2^{c*})^T$ basis.
\\\draftnote{dm - say something about abuse of notation and calling mass eigenstates X1 and X2 as well?  Or ignore mixing?}
\end{comment}
%%%%%%%%%%%%%%%%%%%%%

The $U(1)^\prime$ gauge sector consists of the $Z^\prime$ gauge 
boson, with mass $m_{Z^\prime}^2 = 2 \gzp^2( \eta^2 + \eta_c^2)$, 
the $\widetilde Z^\prime$ gaugino, and the hidden Higgsinos 
$\widetilde H, \widetilde H^c$.  The three neutralinos have the mass matrix
\begin{equation}
M = \left( \begin{array}{ccc} 
M^\prime & -\sqrt{2}\gzp\eta & \sqrt{2}\gzp \eta_c  \\
 -\sqrt{2}\gzp \eta  & 0 & \mu_H \\
 \sqrt{2}\gzp \eta_c  & \mu_H & 0 
\end{array} \right) \, , \label{eq:neutralino}
\end{equation}
which can be brought into a diagonal form using a unitary transformation $P$, 
such that $P^\dagger M P = \mathrm{diag}(m_{\chi_1},m_{\chi_2},m_{\chi_3})$.
The $U(1)^\prime$ gauge superfield  mixes kinetically with the MSSM
hypercharge,
\beq
-\mathscr{L}\supset \frac{\kappa}{2} \int d^2\theta B^\alpha Z'_\alpha 
\ + \mathrm{h.c.}\ ,
\eeq 
where $Z'_\alpha$ and $B_\alpha$ are the $U(1)'$ and $U(1)_Y$ supersymmetric gauge
field strengths, respectively, with the mixing parameter $\kappa \ll 1$.

  The full particle content of the hidden sector after the spontaneous breaking of $U(1)'$ consists of the following mass eigenstates: three neutralinos $\chi_i$; three hidden Higgs scalars $h$, $H$, $A$; two Dirac fermions $\Psi_i$; four complex scalars $\Phi_i$; and a massive gauge boson $Z'$.  The lightest Dirac fermion $\Psi_1$ and complex scalar $\Phi_1$ are stable due to their masses and $B$ charge assignments --- they make up the dark matter. All other states either annihilate or decay into Standard Model particles as described in
Sec.~\ref{sec:baryo}.

Now that we have presented the ingredients for the hidden sector states, we make some remarks:
\begin{itemize}
\item We assume that the mass scales of the hidden sector parameters lie 
at the GeV scale (with the exception of the $X$ states).  
For the soft terms, this can be accomplished by assuming that SUSY-breaking is 
suppressed in the hidden sector (see Sec.~\ref{sec:susybusting}).  
However, the SUSY-preserving vector mass terms present a hidden 
$\mu$-problem; we ignore this issue, but in principle this can be 
solved by introducing an additional hidden singlet analogous to the NMSSM.
\item Since $X_{1,2}$ mediates baryon transfer between the visible 
and hidden sectors (described below), IND signatures are more 
favorable if the DM states $(\Phi_1, \Psi_1)$ are mostly aligned 
with the $Y_1$ supermultiplet.  However, nonzero mixing with $Y_2$ is
induced by SUSY-breaking and hidden Higgs vevs resulting in a DM-$Z^\prime$ 
coupling that is essential for annihilation of the symmetric DM density.
\item We have imposed $B$ as a global symmetry.  Since gravitational 
effects are expected to violate global symmetries, $B$ violation 
could arise through Planck-suppressed operators, potentially leading to DM
particle-antiparticle oscillations that can erase the hidden 
baryon asymmetry~\cite{Cohen:2009fz,Buckley:2011ye,
Cirelli:2011ac,Tulin:2012re}.  
In our SUSY framework, these effects are forbidden by the $\mathbb{Z}_4^R$
extension of $R$-parity. For example, the $B=-1$ operators 
$W\sim MY_1^2,~Y_1 Y_2 H^c$ are allowed by $U(1)'$ and can lead to 
DM oscillations, but they are not invariant under $R$-parity. If $\mathbb{Z}_4^R$ descends from an anomaly-free
gauge symmetry,
%\footnote{A discrete remnant of a gauge symmetry must remain anomaly free. In particular $\mathbb{Z}_4^R$ is anomaly free since we extended the usual, 
%anomaly free~\cite{Ibanez:1991hv}, $\mathbb{Z}_2^R$ only by adding vector-like states.}
such as $U(1)_{B-L}$ spontaneously broken by two units, 
it cannot be violated by gravity~\cite{Ibanez:1991hv,Krauss:1988zc} and these operators are forbidden. 
Thus there exists a consistent embedding of $\mathbb{Z}_4^R$ in a gauge symmetry that excludes the
Majorana mass terms for $Y_{1,2}$ that could erase the hidden asymmetry by  oscillations.

%In our SUSY framework, these effects are forbidden by our $\mathbb{Z}_4^R$
%extension of $R$-parity provided this symmetry descends from a genuine 
%gauge symmetry~\cite{Krauss:1988zc}, a notable candidate being $U(1)_{B-L}$. 
%As long as $U(1)_{B-L}$ is spontaneously broken by more than one unit, 
%Majorana mass terms for $Y_{1,2}$ are forbidden and  
%the hidden antibaryon asymmetry cannot be erased by oscillations.
\end{itemize}

\subsection{Baryon transfer}

Baryon number is transferred between the hidden and visible sectors through superpotential terms $W_{\rm trans}$. The hidden baryon states $X_{1,2}$ are coupled to the operator $U^c_i D^c_j D^c_k$, where $U^c_i, D_j^c$ are the usual $SU(2)_L$-singlet quark superfields ($i,j,k$ label generation).  We focus on the case involving light quarks ($U^c \equiv U_1^c$, $D^c\equiv D_1^c$, $S^c \equiv D_2^c$), corresponding to the ``hyperon portal''~\cite{Dimopoulos:1987rk}:
\beq
W_{\rm trans} =\sum_{a=1,2} \frac{\lambda_{a}}{M} \epsilon_{\alpha\beta\gamma} X_{a} {U}_{\alpha}^c D_{\beta}^c S_{\gamma}^c \; , \label{trans}
\eeq
with $SU(3)_C$ indices $\alpha,\beta,\gamma$ and nonrenormalizable couplings $\lambda_{1,2}/M$.  Although hylogenesis is viable for any generational structure, Eq.~\eqref{trans} is the most interesting case for IND signatures.  In contrast to non-SUSY hylogenesis models, the ``neutron portal'' coupling $X_{1,2} U^c D^c D^c$ vanishes by antisymmetry.  SUSY hylogenesis therefore favors IND involving $K$ final states, rather than $\pi, \eta$ final states allowed in generic non-SUSY models.  

The simplest possibility to generate the nonrenormalizable coupling in Eq.~\eqref{trans} is to introduce a vector-like colour triplet supermultiplet $P$ with global charges $B=-2/3$ and $R=1$.  There are three cases to consider:\footnote{We consider the interactions of cases II and III separately, although in general both may arise simultaneously. The simultaneous presence of both sets of couplings leads to strangeness-violating interactions that may be constrained by flavor violation constraints that we do not consider here.}
\beq
W_{\rm trans} = \left\{ \begin{array}{ll} \blambda_{1,2} X_{1,2} P_\alpha U^c_\alpha  + \bdlambda\, \epsilon_{\alpha\beta\gamma} P^c_\alpha S^c_\beta D^c_\gamma + \mu_P P_\alpha P^c_\alpha & \quad \textrm{(case I)}\\
\blambda_{1,2} X_{1,2} P_\alpha D^c_\alpha + \bdlambda \,\epsilon_{\alpha\beta\gamma} P^c_\alpha  U^c_\beta S^c_\gamma  + \mu_P P_\alpha P^c_\alpha & \quad \textrm{(case II)} \\
\blambda_{1,2} X_{1,2} P_\alpha S^c_\alpha + \bdlambda \,\epsilon_{\alpha\beta\gamma} P^c_\alpha D^c_\beta U^c_\gamma  + \mu_P P_\alpha P^c_\alpha & \quad \textrm{(case III)} \end{array} \right. \label{UVmodel}
\eeq
The $SU(3)_C \times SU(2)_L \times U(1)_Y$ quantum numbers for $P$ are $(3,1,2/3)$ for case I (up-type), and $(3,1,-1/3)$ for cases II and III (down-type).  In all cases $P^c$ carries the opposite charges.  The choice between the cases in Eq.~\eqref{UVmodel} makes little difference for hylogenesis cosmology.  However, the different cases affect the IND signals, manifested in the ratio of the rates of $p \to K^+$ to $n \to K^0$ channels, discussed in Sec.~\ref{sec:pheno}.

Integrating out $P$ and $P^c$ at the supersymmetric level generates the superpotential operator of Eq.~\eqref{trans} with $\lambda_{a}/M \equiv \blambda_{a} \bdlambda/\mu_P$ together with the (higher-order) K\"{a}hler potential term (for case I, with similar operators for cases II and III)
\beq
K \supset \frac{|\bdlambda|^2}{|\mu_P|^2}\left[(D^{c\dagger}D^c)(S^{c\dagger}S^c) - (S^{c\dagger}D^c)(D^{c\dagger}S^c)\right] \ .  
\label{newkahler}
\eeq
Including supersymmetry breaking leads to additional operators.  In particular, the holomorphic soft scalar coupling $b_P\tilde{P}\tilde{P}^c$ (or a squark-gaugino loop with a gaugino mass insertion) gives rise to the four-fermion operator $X u^c_R d^c_R s^c_R$ that plays a central role in IND.

\section{Hylogenesis cosmology\label{sec:baryo}}

  We turn next to a study of the early Universe dynamics of our supersymmetric model of hylogenesis.  To summarize the main ingredients:
\begin{itemize}
\item We assume that the Universe is dominated at early times by a long-lived nonrelativisitic state $\varphi$ (e.g., an oscillating modulus field), which decays and reheats the Universe before the onset of Big Bang nucleosynthesis~(BBN)~\cite{Moroi:1999zb}.
\item Nonthermal $\cp$-symmetric densities of $X_1$ and $\widetilde X_1$ states are populated through $\varphi$ decays.  Depending on their specific origin, the scalar or the fermion can be created preferentially~\cite{Dine:1995uk}.  $\cp$-violating decays of $X_1$ and $\widetilde X_1$ generate equal-and-opposite asymmetries in quarks and hidden sector baryons ($\Psi, \Phi$), while the total baryon number is conserved.\footnote{We neglect $\cp$-violating decays of $X_2$ and $\widetilde X_2$, which in principle can also contribute to $B$ asymmetries.}
\item A hidden $U(1)^\prime$ gauge sector allows for cascade decays of heavier $B=-1/2$ states ($\Phi_2, \Psi_2$, {\it etc.}) into the lightest states $\Phi \equiv \Phi_1$ and $\Psi \equiv \Psi_1$ that are DM.  Both states can be stable provided the condition $|m_{\Psi}-m_{\Phi}| < m_p+m_e$ is met.  Also, the symmetric DM densities annihilate efficiently through the light $Z^\prime$, which decays to SM states via kinetic mixing with hypercharge.
\end{itemize}
Below, we first compute the $\cp$ asymmetries for $X_1$ and $\widetilde X_1$ decays. Second, we ensure that the successful predictions of BBN are not modified by hidden sector decays into SM particles.  Third, we solve the system of Boltzmann equations for hylogenesis, incorporating all of the aforementioned ingredients, to compute the baryon asymmetries.  There are significant differences compared to nonsupersymmetric hylogenesis~\cite{Davoudiasl:2010am}; in particular, the DM masses $m_{\Phi}$, $m_{\Psi}$ and the ratio of $\Phi$ to $\Psi$ states can be different, with implications for IND phenomenology.

\subsection{$\cp$-violating asymmetries\label{sec:cpviolasym}}

\begin{figure}[ttt]
\begin{center}
\vspace{1cm}
\if\withFigures1
        \includegraphics[width=0.85\textwidth]{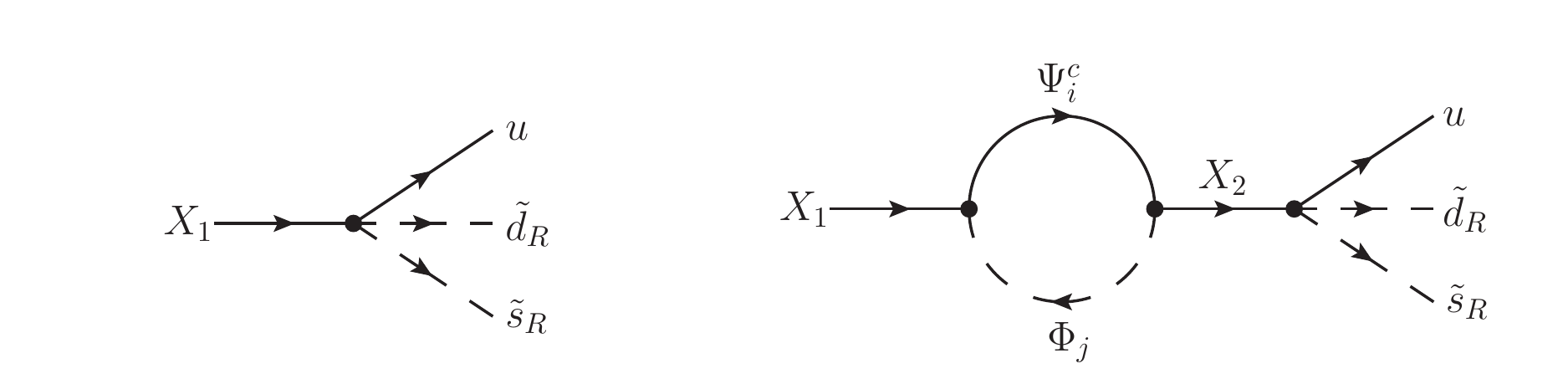}
\fi
\end{center}
\caption{Representative diagrams contributing to 
$X_1\rightarrow q_i\tilde{q}_{Rj}\tilde{q}_{Rk}$ decays which are responsible for 
the generation of the baryon asymmetry.}
\label{fig:massbasishylo}
\end{figure}

Visible and hidden $B$ asymmetries are produced by $\cp$ violation in the partial decay widths of $X_1$
\beq
X_1 \rightarrow u_R\tilde{d}_R\tilde{s}_R + d_R \tilde{s}_R \tilde{u}_R + s_R \tilde{u}_R\tilde{d}_R \; , \quad
X_1 \rightarrow \bar{\Psi}_i \Phi^*_j \; ,
\eeq
due to interference between tree-level and one-loop amplitudes, shown in Fig.~\ref{fig:massbasishylo}.  The corresponding $\cp$ asymmetry is
\begin{subequations} \label{epsilon}
\begin{align}
\epsilon_X \: &\equiv\: \frac{1}{~\Gamma_{X_1}} \Big[ \Gamma(X_1\rightarrow u_R \tilde{d}_R\tilde{s}_R)
- \Gamma(\bar{X}_1\rightarrow \bar{u}_R \tilde{d}_R^*\tilde{s}_R^*) + \textrm{perms.} \Big] \\
&= \: \frac{3\left[\textnormal{Im}(\lambda_1^*\zeta_1\zeta_2^*\lambda_2)m_{X_1}
+\textnormal{Im}(\lambda_1^*\bar{\zeta}_1^*\bar{\zeta}_2\lambda_2)m_{X_2}\right]
m_{X_1}^3}
{64\pi^3 M^2 (m_{X_2}^2-m_{X_1}^2)(|\zeta_1|^2+|\bar{\zeta}_1|^2)}.
\end{align}
\end{subequations}
We assume that $\Gamma_{X_1}$ is dominated by the two-body decay to $\bar{\Psi}_i \Phi_j^*$ final states.\footnote{In what follows, flavor indices $i,j$ for hidden sector states are implicitly summed over in final states.}  For $\epsilon_X > 0$, a positive net $B$ asymmetry is generated in the visible sector.  By $\cpt$ invariance, the decay rates for $X_1$ and $\bar{X}_1$ are equal, and so an equal-and-opposite (negative) $B$ asymmetry is generated in the hidden sector.  Additional contributions to $\epsilon_X$ from $X_1 \rightarrow u_R d_R s_R$, arising through SUSY-breaking or at one-loop, may be subleading provided squark decays are kinematically available.

%%%%%%%%%%%%%%%%%%%%%%%%%%%%%%%%%%%%%%%%%%%%%%%%%%%%%%%%%%%%%%%%%%%%%%
%%%%%%%%%%%%%%%%%%%%%%%%%%%%%%%%%%%%%%%%%%%%%%%%%%%%%%%%

The baryon asymmetry can also be generated through 
the decays of the scalar component of the $X_1$ superfield, $\widetilde{X}_1$,
via interference of supersymmetrizations of the 
diagrams in Fig.~\ref{fig:massbasishylo}.
In the supersymmetric limit, the $\cp$-asymmetry due to $\widetilde{X}_1$ is equal to Eq.~\eqref{epsilon}. 
However, $\widetilde X_1$ decay can populate preferentially $\Psi$ or $\Phi$, due to the different hidden sector decay rates
\beq
\Gamma(\widetilde{X}_1 \rightarrow\Phi_i^* \Phi_j^*)  
= \frac{|\bar{\zeta}_1|^2}{16\pi}m_{X_1} \, , \quad
\Gamma(\widetilde{X}_1 \rightarrow\bar\Psi_i \bar\Psi_j)  
= \frac{|\zeta_1|^2}{16\pi}m_{X_1}.
\eeq
For $X_1$ decays, the primordial ratio 
\beq
r\equiv n_\Psi/n_\Phi
\eeq
of charge densities $n_{\Psi,\Phi}$ is equal to unity.  However, $\widetilde X_1$ decays can deviate from $r = 1$ for $|\zeta_1|\ne |\bar \zeta_1|$.  As we discuss below, IND signals can be significantly enhanced if the heavier state is overpopulated compared to the lighter state ({\it e.g.} $r\gg 1$ for $m_\Psi > m_\Phi$).

\begin{figure}[ttt]
\begin{center}
\if\withFigures1
        \includegraphics{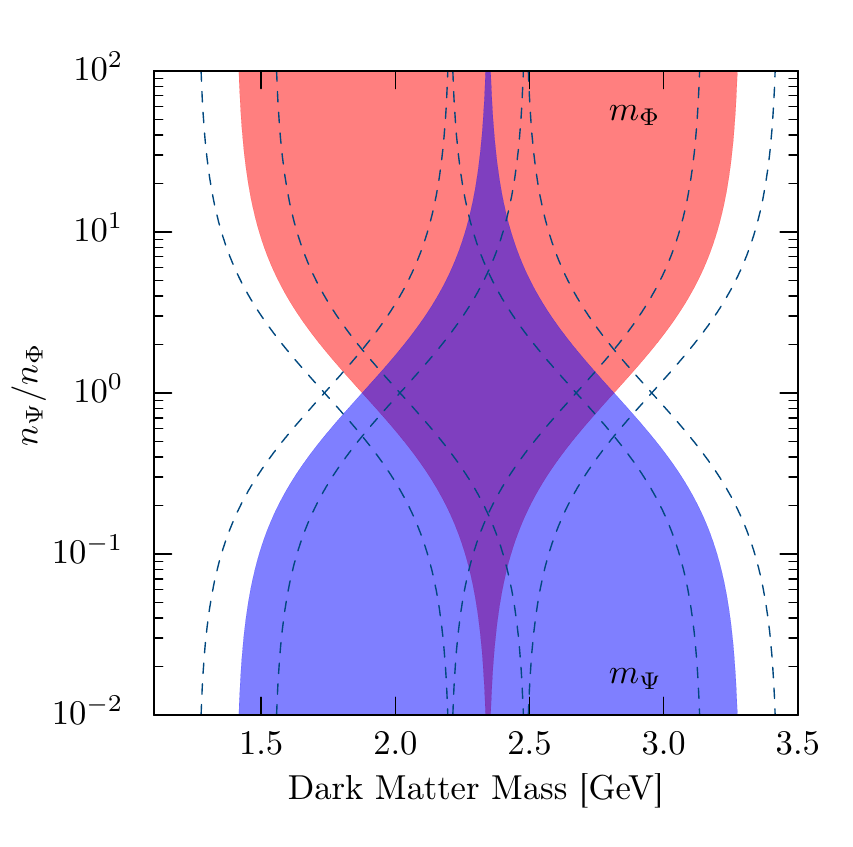}
\fi
\end{center}
\caption{(color online) Allowed masses for the scalar $\Phi$ and fermion $\Psi$
components of dark matter. For a fixed value of $n_\Psi/n_\Phi$, 
the shaded region shows the entire mass range of $\Psi$ (blue) and $\Phi$ (red)
that reproduces $\Omega_\mathrm{DM}/\Omega_{\rm b}\approx5$ and satisfies the stability requirement $|m_\Psi - m_\Phi| < m_e + m_p$. Shifting $\Omega_\mathrm{DM}/\Omega_{\rm b}$ by 
$+(-)6\%$ moves the allowed region right (left), as indicated by the
dashed contours.  
}
\label{fig:phasediag}
\end{figure}

The dark matter abundance is given by 
\beq
\frac{\Omega_{\rm DM}}{\Omega_{\rm b}} = \frac{2(m_\Psi r + m_\Phi)}{m_p(1+r)},
\eeq
where we have neglected the contributions from the DM anti-particles. This
is appropriate in the limit of completely asymmetric DM populations.
The allowed DM mass window, including the uncertainty in $\Omega_\mathrm{DM}/\Omega_{\rm b} \approx 5$, is then
\beq
1.3\,\gev\leq m_\Psi,\,m_\Phi \leq 3.4\,\gev . \label{massrange}
\eeq
More specifically, Fig.~\ref{fig:phasediag} shows the allowed mass range for $m_\Psi$ (blue) and $m_{\Phi}$ (red), for a given value of $r$.  In the $r \to 0$ ($\infty$) limit, only $\Phi$ ($\Psi$) is populated and its mass is required to be approximately $5m_p/2$ to explain the DM density; the underpopulated $\Psi$ ($\Phi$) state is constrained within the range of Eq.~\eqref{massrange} by the stability condition $|m_\Psi - m_\Phi| < m_p + m_e$.

%%%%%%%%%%%%%%%%%%%%%%%%%%%%%%%%%%%%%%%%%%%%%%%%%%%%%%%%%%%%%%%%%%%%%%
%%%%%%%%%%%%%%%%%%%%%%%%%%%%%%%%%%%%%%%%%%%%%%%%%%%%%%%%

\subsection{Decays and Annihilations of SUSY States}

Aside from the stable DM states $\Phi_1$ and $\Psi_1$, the hidden sector contains numerous states that decay, 
producing additional SM radiation.  These decays, listed below, must occur
with a lifetime shorter than about one second to avoid conflicts with BBN predictions.

\begin{itemize}
\item The $Z'$ gauge boson decays to SM states via kinetic mixing with the photon, requiring $\kappa \gtrsim 10^{-11}\, (m_{Z^\prime}/\gev)^{-1}$~\cite{Pospelov:2007mp}, while $\kappa\lesssim 10^{-3}$ is consistent with existing limits for $m_{Z^\prime} \sim \gev$~\cite{Pospelov:2008zw,Bjorken:2009mm}.\footnote{We assume a stronger condition $\kappa \gtrsim 10^{-8} g_*  (m_{Z^\prime}/\gev)^{-1} (\TRH/\gev)^{3/2}$ such that the hidden and visible sectors are in kinetic equilibrium at $T<\TRH$.~\cite{Pospelov:2007mp}}
\item For the hidden Higgs states, the heavy $\cp$-even state decays $H \to Z^\prime Z^\prime$, 
while the $\cp$-odd state decays $A \to Z'h$, through $U(1)^\prime$ gauge interactions.  
Since the lighter $\cp$-even state $h$ is necessarily lighter than the $Z'$, it
must decay to Standard Model fermions either via loop-suppressed
processes~\cite{Batell:2009yf} or via $D$-term mixing with the MSSM Higgs~\cite{Chan:2011aa}. For $h$ masses above the two-muon threshold, the mixing process dominates requiring $\kappa \gtrsim 10^{-5}$.
%For small Higgs masses the loop-induced process dominates~\cite{Chan:2011aa}, requiring $\kappa \gtrsim 10^{-4}/\sqrt{\gzp}$.
\item The heavy dark states $\Phi_i$ ($i>1$) and $\Psi_2$ cascade
down to $\Phi_1$ and $\Psi_1$ by emitting $Z'$ and $h$ bosons. 
\item The hidden neutralinos can decay $\chi_i \to \Phi_1 \bar\Psi_1, \Phi_1^* \Psi_1$ provided this channel is open (assumed below).  If this channel is closed, then the lightest state $\chi_1$ is stable, providing an additional DM component, and must annihilate efficiently via the $t$-channel process $\chi_1 \chi_1 \to Z^\prime Z^\prime$.
\end{itemize}
In addition, the lightest supersymmetric particle within the MSSM decays to hidden states through mixing of the hidden and MSSM neutralinos induced by $\kappa$.
This mixing generally has a negligible effect on the mass eigenvalues~\cite{Chan:2011aa}.

The symmetric DM densities of $\Psi_1 \bar\Psi_1$ 
and $\Phi_1 \Phi_1^*$ annihilate to $Z^\prime$ gauge bosons.  
In the case where DM is nearly aligned with the $Y_2$ multiplet, 
the cross sections are given by \cite{Pospelov:2007mp}
\beq
\label{eq:annixsec}
\langle\sigma v 
\rangle_{\Psi\bar{\Psi}\rightarrow Z'Z'} =
\frac{\gzp^4}{16\pi m_\Psi^2}\sqrt{1-m_{Z'}^2/m_\Psi^2} , \quad
\langle\sigma v 
\rangle_{\Phi\Phi^*\rightarrow Z'Z'} =
\frac{\gzp^4}{16\pi m_\Phi^2}\sqrt{1-m_{Z'}^2/m_\Phi^2} \, .
\eeq
To have $\langle \sigma v \rangle \gtrsim 3 \times 10^{-26}$ cm$^3$/s for efficient annihilation, we require $\gzp \gtrsim 0.03$~\cite{Graesser:2011wi}.  

The presence of light hidden neutralinos allows for the chemical
equilibration of baryon number between $\Phi$ and $\Psi$.  The most important 
process is $\Phi_1 \Phi_1 \leftrightarrow \Psi_1 \Psi_1$ which 
transfers the $B$ asymmetry from the heavier DM state to the lighter state.  
This effect is phenomenologically important for IND, potentially 
quenching the more energetic down-scattering IND processes.  
The transfer arises from the supersymmetrization of the hidden 
gauge and Yukawa interactions, which, in the mass basis, takes the form
\begin{align}
\mathscr{L}  \supset & \, \Phi_i \overline{\Psi}_j 
\left[\left(-\sqrt{2}\gzp Z_{i4}^* V_{j2} P_{1k} - \gamma Z_{i1}^* V_{j2} P_{2k} - \bar{\gamma} Z_{i2}^* V_{j1} P_{3k}\right)P_L \notag \right.\\ 
  + & \, \left.\left(\sqrt{2}\gzp Z_{i2}^* U_{j2} P_{1k}^* -\gamma^* Z_{i4}^* U_{j1} P_{2k}^* - \bar{\gamma}^* Z_{i3}^* U_{j2}^* P_{3k}^*\right)P_R\right] \chi_k 
+ \textrm{h.c.}
\end{align}
In the limit where the dark matter is mostly aligned with the 
$Y_2$ supermultiplet, the interaction simplifies to 
\beq
\mathscr{L}\supset \sqrt{2}\gzp\Phi_1\overline{\Psi}_1\left(a P_L + b P_R \right)\chi_1
+\textnormal{ h.c.},
\eeq
where $a = -\sqrt{2}\gzp Z^*_{14}$, $b = \sqrt{2}\gzp  Z^*_{12}$,
and $\chi_1$ is the $U(1)'$ gaugino. Note that even if the mixing
due to $U(1)'$ breaking can be neglected, the scalars $\widetilde{Y}_2$
and $\widetilde{Y}^{c*}_2$ will still mix via the soft $b$-term. 
For $m_\Phi \geq m_\Psi$ the $s$-wave contribution to the thermalized 
$\Phi_1\Phi_1\rightarrow\Psi_1\Psi_1$ cross section for this
interaction takes the form
\begin{align}
\langle \sigma v\rangle_{\Phi\Phi\rightarrow\Psi\Psi} &= \frac{1}{8\pi(M^2+m_\Phi^2-m_\Psi^2)^2} \sqrt{1-\frac{m_\Psi^2}{m_\Phi^2}}\times \nnmb\\
 & \times \left(
2m^2_\Phi \right[(|a|^4+|b|^4)m_\chi^2 
+ (|a|^2+|b|^2)(ab^*+a^*b) m_\chi m_\Psi + 2|a|^2 |b|^2 m_\Psi^2\left] \right. \nnmb\\
 & \left. - m_\Psi^2 |(a^2 + b^2) m_\chi + 2abm_\Psi|^2
\right).
\end{align} 
In our numerical calculations we use $a=b=\gzp$ for simplicity. In this case
the transfer cross section reduces to 
\beq
\langle\sigma v
\rangle_{\Phi\Phi\rightarrow\Psi\Psi}
=
\gzp^4 \frac{(m_\chi+m_\Psi)^2}{2\pi(m_\chi^2+m_\Phi^2-m_\Psi^2)^2}
\left(1-\frac{m_\Psi^2}{m_\Phi^2}\right)^{3/2}.
\eeq
The cross section for the reverse process $\Psi_1\Psi_1\rightarrow\Phi_1\Phi_1$ 
is related to $\Phi_1\Phi_1\rightarrow\Psi_1\Psi_1$ by the detailed balance condition
\beq
\langle \sigma v\rangle_{\Phi\Phi\rightarrow\Psi\Psi} = 
 \left(n_\Psi^\mathrm{eq}/n_\Phi^\mathrm{eq}\right)^2
\langle \sigma v\rangle_{\Psi\Psi\rightarrow\Phi\Phi},
\eeq
where the equilibrium distributions $n_i^\mathrm{eq}$ are given
in Eq.~\eqref{eq:mbdist}. 
We discuss depletion of the heavier DM state in more detail below.  
Before doing so, however, let us mention that the transfer process
is not generic, and may be absent in other constructions.
In particular, this is true of the alternate Higgs portal model
presented in Appendix~\ref{sec:hportal}.

\subsection{Boltzmann equations}

\begin{figure}[ttt]
\begin{center}
\vspace{1cm}
\if\withFigures1
        \includegraphics{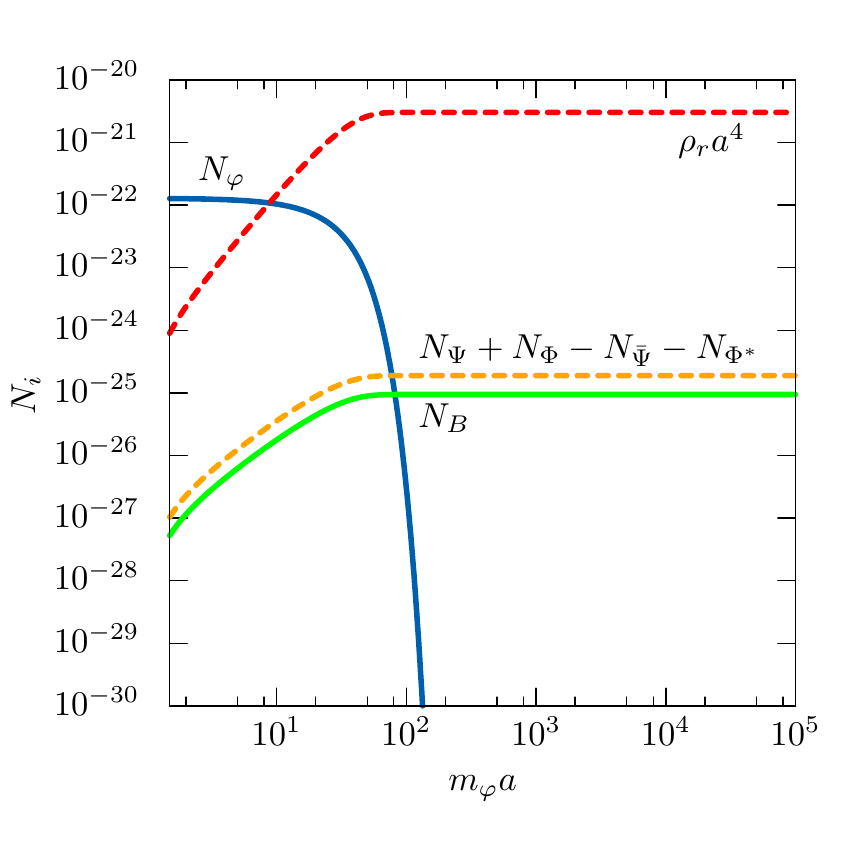}
\fi
\end{center}
\caption{(color online) 
Solutions to the reheating Eqs. 
(\ref{eq:modBoltz},~\ref{eq:entBoltz},~\ref{eq:barBoltz}) and
DM production, described by Eq.~\eqref{eq:darkboltz}. Here
$N_\varphi = \rho_\varphi a^3/m_\varphi$ and $N_i = n_i a^3$ for
$i=B,\,\Psi,\,\Phi,\,\bar{\Psi},\,\Phi^*$.
}

\label{fig:ReheatSol}
\end{figure}

The generation of the visible and hidden $B$ asymmetries 
during reheating is described by a system of Boltzmann equations:
\begin{subequations}
\label{eq:boltz}
\begin{align}
\label{eq:modBoltz}
\dot{\rho}_\varphi = & - 3H\rho_\varphi - \Gamma_\varphi\rho_\varphi \\ 
\label{eq:entBoltz}
\dot{s} = & -3 H s + \frac{\Gamma_\varphi}{T}\rho_\varphi\\
\label{eq:barBoltz}
\dot{n}_B  = &  - 3 H n_B + (\epsilon_X \mathcal{N}_X 
+ \epsilon_{\widetilde{X}} \mathcal{N}_{\widetilde{X}}) 
\frac{\Gamma_\varphi \rho_\varphi}{m_\varphi}.
\end{align}
\end{subequations}
Here, $\rho_\varphi$ is the energy density of the modulus field $\varphi$, $s$ is the entropy density, and $n_B$ is the visible $B$ charge density (the hidden $B$ asymmetry is $-n_B$).  Also, $\mathcal{N}_{X,\,\widetilde{X}}$ is the average number of $X_1$ or its superpartner produced per modulus decay, while $\epsilon_{X,\,\widetilde{X}}$
is the $\cp$-asymmetry from $X_1,\widetilde{X}_1$
decay. In the supersymmetric limit $\epsilon_X=\epsilon_{\widetilde{X}}$.
The modulus decay rate $\Gamma_\varphi$ determines the reheat temperature
\beq
\TRH = \left[\frac{45}{4\pi^3 g_*(\TRH)}\right]^{1/4}\sqrt{\Gamma_\varphi \Mpl}.
\eeq
The total modulus decay rate is given by~\cite{Moroi:1999zb,Allahverdi:2010im, Acharya:2010af} 
\beq
\Gamma_\varphi = \frac{m_\varphi^3}{4\pi\Lambda^2},
\eeq
where we take $\Lambda = 2.43\times10^{18}~\gev$ to be the reduced Planck constant.  Along with the Friedmann equation $H^2 = (8\pi G/3)(\rho_\varphi + \rho_r)$,
where $\rho_r = (\pi^2/30)g_* T^4$, Eqs.~\eqref{eq:boltz} 
form a closed set and
can be solved using the method of Refs.
\cite{Chung:1998rq,Giudice:2000ex}. Here $g_*$ is the energy density number of
relativistic degrees of freedom. Instead of entropy, one can also
solve for radiation density. We take $m_\varphi=2000~\tev$ 
(corresponding to $\TRH \approx 270~\mev$), 
$\mathcal{N}_X=\mathcal{N}_{\widetilde{X}}=1$ and 
$\epsilon_X=\epsilon_{\widetilde{X}} = 3.68\times10^{-4}$. 
This decay asymmetry can be generated for example by 
taking $|\lambda_a|\approx 1$, 
$|\zeta_a|\approx 0.1$,\footnote{The magnitude of the coupling constants 
is chosen to be consistent with hidden sector 
SUSY-breaking as discussed in Sec.~\ref{sec:susybusting}.}  
maximal $\cp$-violating phase and 
$M\approx m_{X_1} \approx m_{X_2}/3 \approx 1\,\tev$.
These parameters reproduce the observed baryon asymmetry 
$\eta_B=n_B/s\approx8.9\times10^{-11}$. 
Numerical solutions to the reheating equations
for these parameters are shown in Fig.~\ref{fig:ReheatSol}. 
The modulus field $\varphi$ decays into radiation and the
heavy states $X$, which immediately decay asymmetrically in the
visible and hidden sectors, generating the baryon asymmetry 
and the dark matter abundance.
%%%%%%%%%%%%%%%%%%%%%%%%%%%%%%%%%%%%%%%%%%%%%%%%
%%%%%%%%%%%%%%%%%%%%%%%%%%%%%%%%%%%%%%%%%%%%%%%%%%%%

The production
of dark matter and its dynamics are described by a system of four Boltzmann 
equations for $\Psi_1$, $\Phi_1$ and their antiparticles
\footnote{We assume that the heavier dark states $\Psi_2$ and $\Phi_i$, $i>1$, decay
to $\Psi_1$ and $\Phi_1$ sufficiently fast.} which take the form
\beq
\label{eq:darkboltz}
\dot{n}_i = -3 H n_i + C_i + (
\mathcal{N}_{i,X}\mathcal{N}_X+\mathcal{N}_{i,\widetilde{X}}\mathcal{N}_{\widetilde{X}})
\frac{\Gamma_\varphi \rho_\varphi}{m_\varphi},
\eeq
for $i=\Psi_1,\,\bar{\Psi}_1,\,\Phi_1,\Phi^*_1$. Here $\mathcal{N}_{i,X}$ is the
average number of species $i$ produced per $X$ decay. For $i=\Psi$ 
(we drop the subscript $1$ from hereon) we have 
\beq
\mathcal{N}_\Psi\equiv\mathcal{N}_{\Psi,X}+ \mathcal{N}_{\Psi,\widetilde{X}}
= \frac{\Gamma(\bar{X}\rightarrow\Psi\Phi)
+2\Gamma(\widetilde{X}^*\rightarrow\Psi\Psi)}{\Gamma_X}.
\label{eq:avenumPsi}
\eeq
Similar definitions hold for $\bar{\Psi}$, $\Phi$ and $\Phi^*$, so that 
\beq
\mathcal{N}_\Psi + \mathcal{N}_\Phi - 
\mathcal{N}_{\bar{\Psi}} - \mathcal{N}_{\Phi^*}
= 2 \epsilon_X + 2\epsilon_{\widetilde{X}}.
\eeq
 The last term on the right hand side of Eq.~\eqref{eq:darkboltz}
describes the production of the species $i$ through modulus decay into $X$ which
promptly decays into $i$. The quadratic collision terms $C_i$ describe the 
particle-antiparticle annihilations as well as the transfer reaction 
$\Psi\Psi\leftrightarrow\Phi\Phi$, required by supersymmetry.
The collision terms for $i=\Psi,\,\Phi$  are
\begin{align}
C_\Psi =& -\langle\sigma v 
\rangle_{\Psi\bar{\Psi}\rightarrow Z'Z'} 
\big[n_\Psi n_{\bar{\Psi}} - (n_\Psi^\mathrm{eq})^2 
\big]
- \langle\sigma v
\rangle_{\Psi\Psi\rightarrow\Phi\Phi}
\big[
n_\Psi^2-\left(n_\Psi^\mathrm{eq}/n_\Phi^\mathrm{eq}\right)^2 n_\Phi^2
\big]\\
C_\Phi =& -\langle\sigma v 
\rangle_{\Phi\Phi^*\rightarrow Z'Z'} 
\big[n_\Phi n_{\Phi^*} - (n_\Phi^\mathrm{eq})^2\big]
- \langle\sigma v
\rangle_{\Phi\Phi\rightarrow\Psi\Psi}
\big[
n_\Phi^2-\left(n_\Phi^\mathrm{eq}/n_\Psi^\mathrm{eq}\right)^2 n_\Psi^2
\big],
\end{align}
where 
\beq
n_i^\mathrm{eq} = \frac{g_i}{2\pi^2} T m_i^2 K_2(m_i/T)
\label{eq:mbdist}
\eeq
is the Maxwell-Boltzmann equilibrium number density for a particle of mass $m_i$ with
$g_i$ internal degrees of freedom. The collision terms for the antiparticles are
identical, with the replacements $\Psi\rightarrow\bar{\Psi}$ 
and $\Phi\rightarrow\Phi^*$. 

%%%%%%%%%%%%%%%%%%%%%%%%%%%%%%%%%%%%%%%%%%%%%%%%
\begin{figure}[ttt]
\begin{center}
\vspace{1cm}
\if\withFigures1
        \includegraphics[scale=0.94]{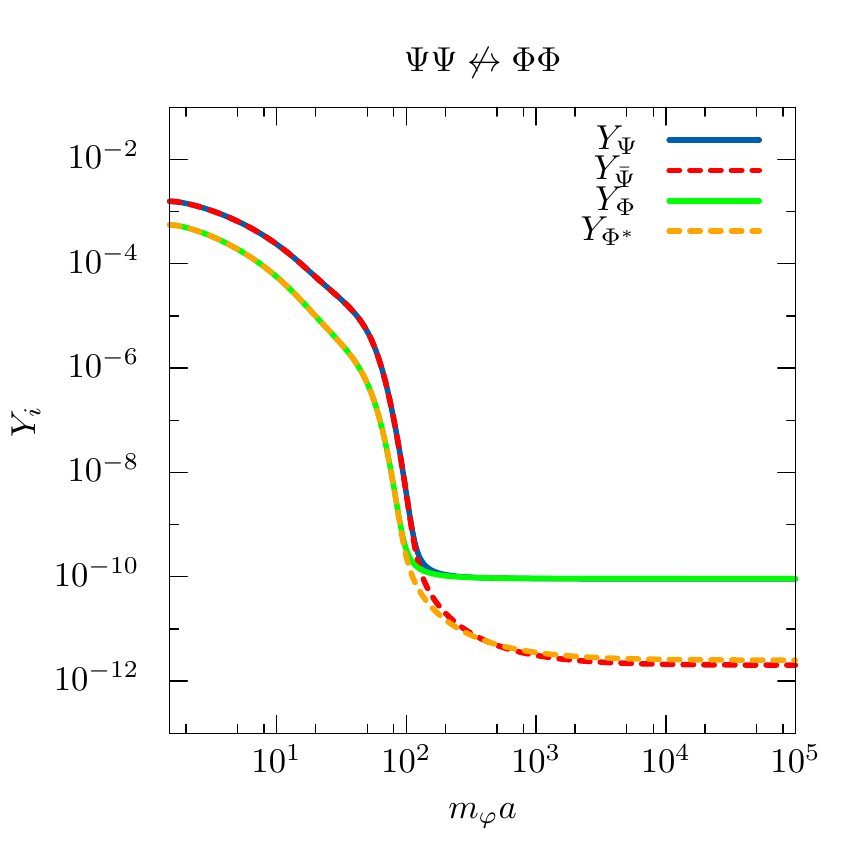}
        \includegraphics[scale=0.94]{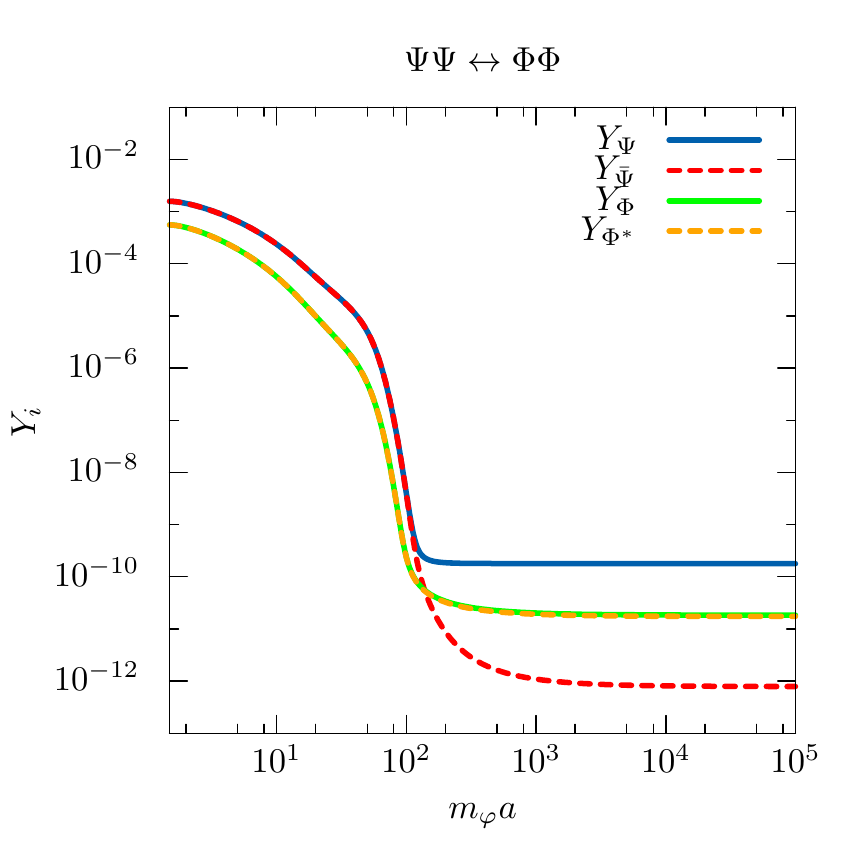}
\fi
\end{center}
\caption{(color online)   
Solution to the Boltzmann equations for the yields $Y_i = n_i/s$
as a function of the scale factor $a$. The plot on the left
shows the evolution for the case when the transfer reaction
$\Phi\Phi\leftrightarrow\Psi\Psi$ is turned off, while the 
plot on the right shows the outcome when it is active. The
transfer drives the dark matter population into lighter 
state, $\Psi$ in this case. 
The DM (anti-DM) abundance is indicated by solid (dashed) lines, with dark (light) lines referring to the fermion (scalar) component.
The parameters used are
described in the text.
}
\label{fig:BoltzmannSol}
\end{figure}
%%%%%%%%%%%%%%%%%%%%%%%%%%%%%%%%%%%%%%%%%%%%%%%%%%%%

The solutions to the Boltzmann equations for the yields
$Y_i = n_i/s$ are shown in Fig.~\ref{fig:BoltzmannSol} for
$m_\Psi=1.9\,\gev$, $m_\Phi=2.2\,\gev$, $m_\chi=5\,\gev$, $\gzp=0.05$. 
We consider two cases. In the plot on the left, we show
the limit where $\langle\sigma v\rangle_{\Phi\Phi\leftrightarrow\Psi\Psi} = 0$;
this can occur when the rate is mixing-suppressed, for a heavy gaugino, or
within models with alternative symmetric annihilation mechanisms (see Appendix). 
With the transfer turned off, the 
scalar and fermion DM sectors are decoupled. The resulting
DM abundances are determined by the $X$ and $\widetilde{X}$ 
decay asymmetries. In this limit, the dark sector reduces to two 
independent copies of the standard asymmetric DM scenario.
We show the case where $\Psi$ and $\Phi$ are populated 
equally by the $\widetilde{X}$ decays, but, in general, the asymmetries
can be different for the scalar and fermion DM, as  
discussed in Sec.~\ref{sec:cpviolasym}. 

In the plot on the right we show the result
when the transfer is efficient, driving the dark matter population into
the lighter state $\Psi$. Since the asymmetry is also transferred into 
the lighter state, the $\Psi\bar{\Psi}$ annihilation rate is enhanced, 
resulting in a highly asymmetric
final abundance. The heavier state, on the other hand, freezes out
with nearly equal abundances of particle and anti-particle, which are 
about an order of magnitude smaller than that of $\Psi$. The
transfer reaction does not affect the production of the net
hidden sector baryon number 
 $n_\Psi+n_\Phi-n_{\bar{\Psi}}-n_{\Phi^*}$.
Its evolution is shown in Fig.~\ref{fig:ReheatSol}. Note that
\beq
n_\Psi+n_\Phi-n_{\bar{\Psi}}-n_{\Phi^*} = 2 n_B
\eeq
as required by $B$ conservation in hylogenesis.

%%%%%%%%%%%%%%%%%%%%%%%%%%%%%%%%%%%%%%%%%%%%%%%%
\begin{figure}[ttt]
\begin{center}
\vspace{1cm}
\if\withFigures1
        \includegraphics{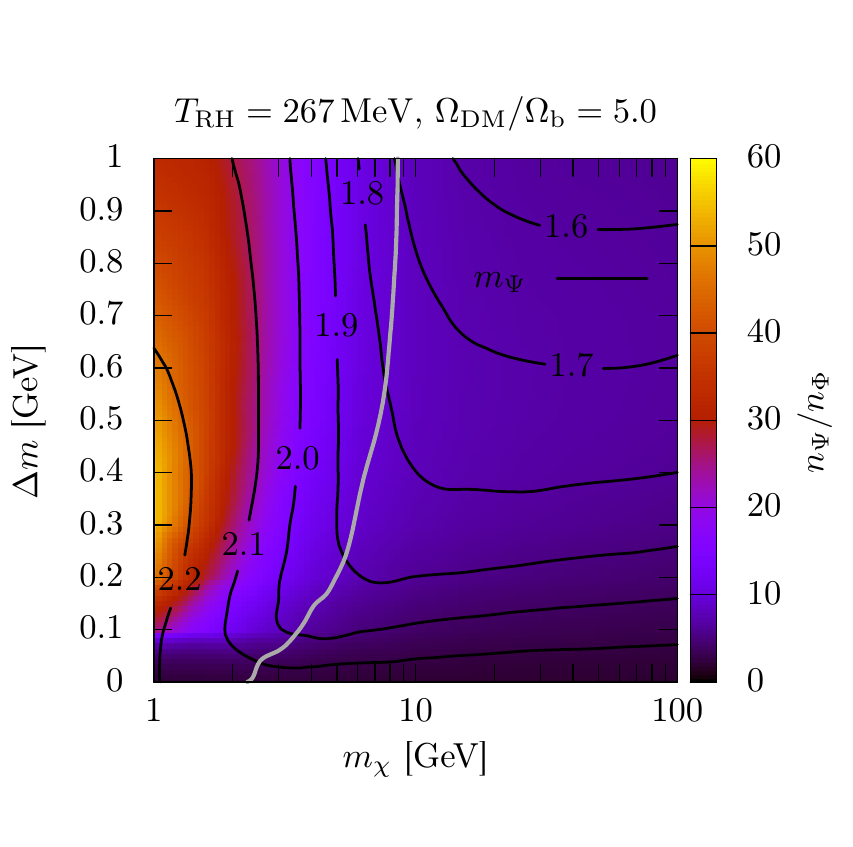}
\fi
\end{center}
\caption{(color online) 
The ratio $n_\Psi/n_\Phi$ for the allowed range of mass
splittings $\Delta m = m_\Phi - m_\Psi$ and relevant
values of the hidden gaugino mass $m_\chi$. At each point 
in the plane the DM abundance is fixed to be  
$\Omega_{\mathrm{DM}}/\Omega_{\mathrm b} = 5.0$. Contours
of constant $m_\Psi$ (in $\gev$) are also shown. The gray
contour shows the CMB constraint for DM annihilations from
Ref.~\cite{Lin:2011gj}. Points to the right of this line
are excluded.
}
\label{fig:paramSpace}
\end{figure}
%%%%%%%%%%%%%%%%%%%%%%%%%%%%%%%%%%%%%%%%%%%%%%%%%%%%

  The ratio of the abundances of $\Psi$ to $\Phi$ is important
for IND. We study the effect of varying the mass splitting
$\Delta m = m_\Phi-m_\Psi$ and the gaugino mass on $n_\Psi/n_\Phi$ in
Fig.~\ref{fig:paramSpace}. For each point in the parameter space, we
solve the reheating and DM production equations and plot the final
value of $n_\Psi/n_\Phi$. The reheat temperature, asymmetry and 
gauge coupling strength are the same as for Fig.~\ref{fig:BoltzmannSol}.
Setting the DM abundance to the observed value fixes $m_\Psi$.
Light gauginos and large DM mass
splittings make the transfer more efficient, increasing the abundance
of the lighter state relative to the heavier one. 
For small $\Delta m$ or heavy $m_\chi$ the transfer rate is suppressed.

If the symmetric density does not annihilate efficiently, 
residual annihilations during the CMB era can inject enough
energy to alter the power spectrum. The WMAP7 constraint on
the annihilation rate for Dirac fermions or complex scalars is
given by~\cite{Lin:2011gj}  
\beq
2\frac{\Omega_i \Omega_{\bar{i}}}{\Omega_{\rm DM}^2}\frac{f\langle \sigma v \rangle}{m_i} < \frac{2.42\times10^{-27}\,\rm{cm^3/s}}{\gev},
\eeq
where $i=\Psi,\,\Phi$ and $\Omega_i/\Omega_{\rm DM}$ is the 
fraction of total DM abundance in species $i$.  
This constraint is shown 
in Fig.~\ref{fig:paramSpace} by the gray line (parameter space to the right is excluded).
For the parameters we have chosen, symmetric annihilation is only marginally efficient, 
and transfer processes through the light gaugino help achieve efficient annihilation (hence, the large $m_{\chi}$ region is excluded). 
For larger gauge coupling $\gzp=0.1$, symmetric annihilation is much more efficient, 
and CMB constraints are evaded in the entire parameter region in Fig.~\ref{fig:paramSpace}.

  Finally, let us mention that we have not included any baryon washout
processes in our Boltzmann equations.  For $T_{RH} \ll m_{X_1},\,m_{\tilde{q}}$,
the only such processes that are allowed kinematically are 
$\Psi\Phi\leftrightarrow 3\bar{q}$ and the corresponding crossed diagrams.  
These transitions require the exchanges of massive intermediate 
$P$ and $X$ states to occur. These processes are therefore 
well described by effective operators of the kind 
\beq
\mathscr{L} \sim \frac{1}{\Lambda^3_\mathrm{IND}}uds\Psi \Phi,
\eeq
where the scale $\Lambda_\mathrm{IND}$ is defined in Section~\ref{sec:pheno} (see Eq.~(\ref{INDquark})) and the order of fermion contractions depends on which UV completion is 
used in Eq.~(\ref{UVmodel}). We find that the corresponding cross sections are safely smaller than 
the stringent limits found in Ref.~\cite{Ellwanger:2012yg} provided
$m_{X},\,m_{P}\gtrsim 300\,\gev$.  
For example, we can estimate the cross section for
$\Psi\Phi\leftrightarrow 3\bar{q}$ as
\beq
\langle \sigma v\rangle \sim \left(\frac{1}{4\pi}\right)^3  
\frac{m_{\Psi_1}^4}{\Lambda_\mathrm{IND}^6}
= \left(4\times10^{-21}~\gev^{-2}\right)
\left(
\frac{1~\tev}{\Lambda_\mathrm{IND}}
\right)^6
\left(
\frac{m_{\Psi_1}}{3~\gev}
\right)^4.
\label{eq:washoutEst}
\eeq
The authors of Ref.~\cite{Ellwanger:2012yg} found that 
washout is negligible for 
\beq
m_\mathrm{DM} \langle \sigma v\rangle \lesssim10^{-18}~\gev^{-1},
\eeq
which is easily satisfied by Eq.~(\ref{eq:washoutEst}). Even 
if these processes were important, the $X_{1}$ decay asymmetry 
can be adjusted to compensate, as long as the couplings satisfy 
the conditions imposed by SUSY breaking discussed in the next 
section.
Thus, our omission of baryon washout effects is justified.

\section{Supersymmetry Breaking
\label{sec:susybusting}}

  Our model for asymmetric antibaryonic dark matter typically
requires light hidden scalars with masses of a few GeV to obtain an 
acceptable dark matter abundance.
For such masses to be technically natural, the size of soft supersymmetry
breaking felt by the light states should also be near the GeV scale.
This is much smaller than the minimal scale of supersymmetry breaking
felt in the MSSM sector, which must be close to or above the TeV scale
to be consistent with current experimental bounds.  

  Such a hierarchy between visible and hidden sector soft terms 
can arise if the hidden sector feels supersymmetry breaking more weakly 
than the visible.  We examine here the necessary 
conditions for this to be the case based on the interactions required
for hylogenesis.  We also discuss a few specific mechanisms of 
supersymmetry breaking that can give rise to the required spectrum.
Motivated by the desire for large moduli masses, which are frequently
on the same order as the gravitino mass, we pay particular attention
to anomaly mediation.

\subsection{Minimal Transmission of Supersymmetry Breaking}
 
  The interactions we have put forward in Section~\ref{sec:susy} 
will transmit supersymmetry breaking from the MSSM
to the hidden sector.  Thus, a minimal requirement for small hidden-sector 
soft terms is that these interactions do not themselves
create overly large hidden soft masses.  We begin by studying these
effects.

  The states that we wish to remain light derive from the $Y_{1,2}^{(c)}$
and $H^{(c)}$ chiral multiplets.  These multiplets do not couple
directly to the MSSM, but they are connected indirectly by their
interactions with the $X^{(c)}$ states and $U(1)'\!-\!U(1)_Y$ gauge 
kinetic mixing.  Thus, the $X^{(c)}$ multiplets and the gauge kinetic 
mixing will act as mediators to the hidden states.

  Beginning with the $X$ multiplets, they will feel supersymmetry
breaking from their direct couplings to the quarks and the
triplet $P^{(c)}$ given in Eq.~\eqref{UVmodel}.  The transmission of 
supersymmetry breaking can be seen in the renormalization
group~(RG) equations of the soft scalar squared masses of $X_a^{(c)}$
and $P^{(c)}$ (assuming interactions as in case~I):
\bea
(4\pi)^2\frac{dm_{X_a}^2}{dt} &=&
6|\blambda_a|^2(m_{X_a}^2+m_{U^c}^2+m_P^2+|A_{\blambda_a}|^2)
+ 4|\zeta_a|^2(m_{X_a}^2+2m_Y^2+|A_{\zeta_a}|^2) 
\nnmb\\
(4\pi)^2\frac{dm_{X^c_a}^2}{dt} &=&
 4|\bar{\zeta}_a|^2(m_{X^c_a}^2+2m_{Y^c}^2+|A_{\bar{\zeta}_a}|^2) 
\\
(4\pi)^2\frac{dm_{P}^2}{dt} &=&
\sum_a2|\blambda_a|^2(m_{X_a}^2+m_{U^c}^2+m_P^2)-\frac{32}{3}g_3^2|M_3|^2 + \ldots
\nnmb\\
(4\pi)^2\frac{dm_{P^c}^2}{dt} &=&
4|\bdlambda|^2(m_{S^c}^2+m_{D^c}^2+m_{P^c}^2)-\frac{32}{3}g_3^2|M_3|^2 +\ldots
\nnmb
\eea
where the $A_{i}$ are trilinear soft terms corresponding to superpotential
operators, and we have dropped subleading hypercharge contributions
for $P^{(c)}$.
We see that $P$ and $P^c$ will typically obtain full-strength
(TeV) soft masses from their direct coupling to the gluon multiplet, 
and these will be passed on to $X$ in the course of RG evolution.  
Recall as well that $\blambda_{1,2}$ and $\bdlambda$ must both be reasonably 
large for hylogenesis to work.

  Turning next to the $Y$ multiplets, we find for the $Y_1$
\bea
(4\pi)^2\frac{dm_{Y_1}^2}{dt} &=& 
\sum_a 8|\zeta_a|^2\left(m_{X_a}^2+m_{Y_1}^2+|A_{\zeta_a}|^2\right)
+ 2|\gamma|^2\left(m_H^2+m_{Y_1}^2+m_{Y_2^c}^2+|A_{\gamma}|^2\right) 
\nnmb\\
&&\\
(4\pi)^2\frac{dm_{Y_1^c}^2}{dt} &=& 
\sum_a 8|\bar{\zeta}_a|^2\left(m_{X^c_a}^2+m_{Y^c_1}^2+|A_{\bar{\zeta}_a}|^2\right)
+ 2|\bar{\gamma}|^2\left(m_{H^c}^2+m_{Y_1^c}^2+m_{Y_2}^2+|A_{\bar{\gamma}}|^2\right)
\ . \nnmb
\eea
The $Y_2$ multiplets are also charged under the $U(1)'$ hidden gauge
symmetry, which mixes kinetically with hypercharge with strength
$\kappa$.  This leads to additional contributions 
to the running~\cite{Baumgart:2009tn,Morrissey:2009ur,Kumar:2011nj}.
At leading non-trivial order in $\kappa$, we have
\bea
(4\pi)^2\frac{dm_{Y_2}^2}{dt} &=& 
2|\bar{\gamma}|^2\left(m_{H^c}^2+m_{Y_1^c}^2+m_{Y_2}^2+|A_{\bar{\gamma}}|^2\right)
\nnmb\\
&& -8\gzp^2(|M^\prime|^2+\kappa^2|M_1|^2) + 2\gzp^2 S_{Z^\prime}
-2\kappa\sqrt{\frac{3}{5}}\;g_1\gzp S_Y
\nnmb\\
&&\\
(4\pi)^2\frac{dm_{Y_2^c}^2}{dt} &=& 
2|\gamma|^2\left(m_H^2+m_{Y_1}^2+m_{Y_2^c}^2+|A_{\gamma}|^2\right)
\nnmb\\ 
&& -8\gzp^2(|M^\prime|^2+\kappa^2|M_1|^2) - 2\gzp^2S_{Z'}
+2\kappa\sqrt{\frac{3}{5}}\;g_1\gzp S_Y
\ , \nnmb 
\eea
where $S_{Z'} = \tr(Q'm^2)$, $S_Y = \tr(Ym^2)$, $g_1 = \sqrt{5/3}\,g_Y$,
and $M_1$ is the hypercharge gaugino (Bino) mass.
The RG equations for the soft mass of $H$ ($H^c$) has the same
form as for $Y_2^c$ ($Y_2$) but with signs of the last two ``$S$''
terms reversed.

  In addition to these RG contributions, there is $D$-term mixing
between hypercharge and $U(1)^\prime$.  After electroweak symmetry breaking 
in the MSSM sector, this generates an effective 
Fayet-Iliopoulos~\cite{Baumgart:2009tn,Morrissey:2009ur} term in the hidden
sector of the form
\beq
V \supset 
\frac{\gzp^2}{2}\left(|H|^2+|Y_2|^2-|H^c|^2-|Y_2^c|^2-\xi_{FI}\right)^2 \ ,
\eeq
with $\xi_{FI}= -\kappa(g_Y/2\gzp)v^2\cos(2\beta)$, where $v \approx 174\,\gev$
and $\tan\beta$ is the ratio of MSSM Higgs vevs.
This term can be absorbed by shifting the hidden-sector soft masses
by $m_i^2\to (m_i^2-Q'_i\xi_{FI})$.

  The RG equations we have presented here are valid down to the scale
$m_{\soft}$ where the MSSM (and $X$ and $P^{(c)}$ if their supersymmetric
masses are near $m_{\soft}$) states should be integrated out.  This will
generate additional threshold corrections to the hidden-sector soft masses.
However, these lack the logarithmic enhancement of the RG contributions and
are typically subleading.  Thus, putting these pieces together we can 
make estimates for the minimal natural values of the soft terms in the 
hidden sector.  
In terms of $m_{\soft} \sim M_3 \sim \tev$ and $\Delta t = \ln(\Lambda_*/m_{\soft})$
(where $\Lambda_*$ is the scale of supersymmetry-breaking mediation), we find
\bea
m_{P^{(c)}}&\gtrsim& m_{\soft}\\
m_{X^{(c)}}~&\gtrsim& |\lambda^\prime|\,m_{\soft}\lrf{\sqrt{\Delta t}}{6}\\
m_{Y_1^{(c)}}~&\gtrsim& |\zeta\,\lambda^\prime|\,m_{\soft}\lrf{\sqrt{\Delta t}}{6}^2\\
m_{Y_2^{(c)}},m_{H^{(c)}}~&\gtrsim& \max\left\{
|\gamma\,\zeta\lambda^\prime|\,m_{\soft}\lrf{\sqrt{\Delta t}}{6}^{3/2},
~\kappa\,M_1\lrf{\sqrt{\Delta t}}{6},~
\sqrt{\frac{\kappa g_Y}{2\gzp}}\,v \right\}\ .
\eea
Note that $\sqrt{\Delta{t}}\approx 6$ for $\Lambda_* = M_\text{Pl}$.
We see that the soft masses of the $Y_2$ and $H$ multiplets can be
naturally suppressed relative to the MSSM for relatively small couplings.
For example, choosing $\gamma = \gzp = 0.05$, $\blambda_1 = 1$, 
$\kappa\sim 10^{-4}$ and $\zeta = 0.1$ yields soft masses for
the $Y_2^{(c)}$ and $H^{(c)}$ multiplets below a few GeV.  
Therefore the direct coupling of the MSSM to the hidden sector need not 
induce overly large supersymmetry breaking in the hidden sector.

\subsection{Mediation Mechanisms}

  We consider next a few specific mechanisms to mediate supersymmetry
breaking to the MSSM and the hidden sector that will
produce a mass hierarchy between the two sectors.  Motivated
by our desire for large moduli masses, which in supergravity constructions 
are frequently related closely to the gravitino 
mass~\cite{Kachru:2003aw,Fan:2011ua}, 
the mechanism we will focus on primarily is anomaly mediation.  
However, we will also describe a second example using gauge mediation 
with mediators charged only under the SM gauge groups.

  With anomaly mediated supersymmetry 
breaking~(AMSB)~\cite{Randall:1998uk,Giudice:1998xp}, the leading-order 
gaugino mass in the hidden sector is
\beq
M^\prime = \frac{b^\prime \gzp^2}{(4\pi)^2}\,m_{3/2} \ ,
\eeq
where $b^\prime = -4$ is the one-loop $U(1)^\prime$ beta function coefficient.
A similar expression applies to the MSSM gaugino soft masses,
but with $\gzp^2b^\prime$ replaced by the corresponding factor.
Based on this comparison, we see that a much lighter hidden gaugino
will arise for small values of the hidden gauge coupling~\cite{Feng:2011uf}.
For example, with MSSM gaugino masses in the range of a few hundred GeV,
the hidden gaugino mass will be a few GeV for $\gzp/g_{SM} \sim 0.1$, 
corresponding to $\gzp \sim 0.05$.

  The hidden-sector scalar soft masses will also be parametrically
smaller than those of the MSSM if the corresponding Yukawa couplings are
smaller as well.  The explicit AMSB expressions for $Y_1$ and $Y_2^c$ are
\bea
m_{Y_1}^2 &=& \frac{m_{3/2}^2}{(4\pi)^4}
\left[
\gamma^2(3\gamma^2-4\gzp^2)+6\gamma^2\sum_a\zeta_a^2+6\sum_a\zeta_a^2\blambda_a^2
+4\sum_{a,b}\zeta_a^2\zeta_b^2(2+\delta_{ab})\right] \ ,
\nnmb\\
&&\label{eq:amsby}\\
m_{Y^c_2}^2 &=& \frac{m_{3/2}^2}{(4\pi)^4}\left[
\gamma^2(3\gamma^2-4\gzp^2)+2\gamma^2\sum_a\zeta_a^2-16\gzp^2\right] \ .
\nnmb
\eea
We also have $m_H^2=m_{Y_2^c}^2$, while $m_{H^c}^2=m_{Y_2}^2$ are given by
the same expressions with $\gamma\to \bar{\gamma}$ and 
$\zeta_a\to \bar{\zeta}_a$.  The latter point also applies to $m_{Y_1^c}^2$ 
relative to $m_{Y_1}^2$ but with $\blambda_a \to 0$ as well.
Thus, we find GeV-scale soft masses for $Y_2^{(c)}$ and $H^{(c)}$
(and TeV-scale MSSM soft masses) for the fiducial values 
$\zeta_a \sim 0.1$, $\gamma \sim \gzp \sim 0.05$, and $m_{3/2} \sim 100\,\tev$.
Note that in these expressions we have neglected
kinetic mixing effects which are negligible for $\kappa < 10^{-3} \ll \gzp/g_1$,
as we assume here.

  The result of Eq.~(\ref{eq:amsby}) shows that the 
AMSB scalar squared masses can be positive or negative, depending on
the relative sizes of the gauge and Yukawa couplings.  This feature
creates a severe problem in the MSSM where minimal AMSB produces
tachyonic sleptons.  We assume that one of the many proposed solutions 
to this problem corrects the MSSM soft masses without significantly
altering the soft masses in the hidden 
sector~\cite{Pomarol:1999ie,Chacko:1999am}.
In contrast to the MSSM, negative scalar soft squared masses need not be 
a problem in the hidden sector due to the presence of supersymmetric mass 
terms for all the multiplets.  In particular, the supersymmetric mass 
terms we have included in Eqs.~(\ref{WHS},~\ref{UVmodel}) for the vector-like 
hidden multiplets can generally be chosen so that only the 
$H$ and $H^c$ multiplets develop vevs. %such that $m_{Z'} \sim \gev$.

  Let us mention, however, that supersymmetric mass terms are problematic
in AMSB.  In particular, a fundamental supersymmetric mass term $M_i$
will give rise to a corresponding holomorphic bilinear soft ``$b_i$'' 
term of size $b_i \sim M_im_{3/2}$.  If $b_i \gg  m^2_{\soft},\,|M_i|^2$, 
such a term will destabilize the scalar potential.
To avoid this, we must assume that the supersymmetric mass terms
we have written in Eqs.~(\ref{WHS},\,\ref{UVmodel}) are generated 
in some other way, such as from the vev of a singlet field.\footnote{
Note that we could have $|\mu_P|, \,|\mu_X| \gg m_{3/2}$ without any problems.
In this case, the threshold corrections to the light soft masses from 
integrating out the heavy multiplets precisely cancel their leading
contributions from RG, leading to a zero net one-loop AMSB contribution.}
A full construction of such a remedy lies beyond the
scope of the present work, but we expect that it can be achieved in analogy
to the many similar constructions addressing the corresponding
$\mu\!-\!B\mu$ problem within the MSSM~\cite{Pomarol:1999ie,Chacko:1999am}
or beyond~\cite{Feng:2011uf}.

  A second option for the mediation of supersymmetry breaking
that preserves the MSSM-hidden mass hierarchy is gauge mediation
by messengers charged only under the SM 
gauge groups~\cite{Baumgart:2009tn,Morrissey:2009ur}.  The soft
masses generated in the hidden sector in this case can be deduced from
the RG equations, up to boundary terms at the messenger scale
on the order of $\kappa\,m_{\soft}$, which are safely small.
Unfortunately, the $U(1)^\prime$ gaugino mass generated in this scenario
only appears at very high loop order, and tends to 
be unacceptably small~\cite{Morrissey:2009ur}.
This can be resolved if there are additional gravity-mediated contributions
to all the soft masses on the order of a few GeV.  The gravitino mass
in this case will be on the same order as the hidden states.  
If it is slightly lighter, it may permit the decay $\Psi_1\to \psi_{3/2}+\Phi_1$
(for $m_{\Psi_1}> m_{3/2}+m_{\Phi_1}$).

\section{Phenomenology}

\label{sec:pheno}

\subsection{Induced Nucleon Decay}

Dark matter provides a hidden reservoir of antibaryons.  Although baryon transfer interactions are weak enough that visible baryons and hidden antibaryons are kept out of chemical equilibrium today, they are strong enough to give experimentally detectable signatures of DM-induced nucleon destruction.  In these events, a DM particle scatters inelastically on a nucleon $N=p,n$, producing a DM antiparticle and mesons.   For SUSY models, the simplest IND events are those involving a single kaon,
\beq
\Psi N \to \Phi^* K  \, , \quad \Phi N \to \bar\Psi K \; . \label{INDprocess}
\eeq
We consider only the lightest DM states $\Psi \equiv \Psi_1$ and $\Phi \equiv \Phi_1$; the heavier states are not kinematically accessible provided their mass gap is larger than $(m_N - m_K) \approx 400$ MeV.  Both down-scattering and up-scattering can occur (defined as whether the heavier DM state is in the initial or final state, respectively), but up-scattering is kinematically forbidden for $|m_\Psi - m_\Phi| < m_N -m_K$.  

Assuming the hidden states are not observed, IND events mimic standard nucleon decay events $N \to K \nu$, with an unobserved neutrino $\nu$ (or antineutrino $\bar \nu$).  Nucleon decay searches by the Super-Kamiokande experiment have placed strong limits on the $N$ lifetime $\tau$ for these modes~\cite{Kobayashi:2005pe}:
\beq
\tau(p \to K^+ \nu) > 2.3 \times 10^{33} \; \textrm{years} \,, \quad \tau(n \to K^0_S \nu) > 2.6 \times 10^{32} \; \textrm{years} . \label{SNDlims}
\eeq
However, these bounds do not in general apply to IND, due to the different kinematics.  For $N \to K \nu$, the $K$ has momentum $p_K \approx 340$ MeV.  IND events are typically more energetic: $680 \lesssim p_K \lesssim 1400$ MeV for down-scattering, and $p_K \lesssim 680$ MeV for up-scattering (if allowed).\footnote{For fixed DM masses, IND is either bichromatic or monochromatic, depending on whether up-scattering is allowed or not; the range in $p_K$ corresponds to the allowed mass range $1.4 \lesssim m_{\Phi, \Psi} \lesssim 3.3$ GeV.  If other hidden states $\Psi_{a\ge 2}$ and $\Phi_{b \ge 2}$ are kinematically accessible, the IND spectrum can have additional spectral lines.}  The Super-Kamiokande analysis assumes: for $p \to K^+$, $K^+$ is emitted below \u{C}erenkov threshold in water, corresponding to $p_K \lesssim 550$ MeV; for $p \to K^0_S$, the $K^0_S$ is emitted with $200 < p_K < 500$ MeV.  Therefore, IND is largely unconstrained by standard nucleon decay searches.  The limits in Eq.~\eqref{SNDlims} only constrain up-scattering IND in a subset of parameter space, whereas down-scattering provides typically the dominant contribution to the total IND rate~\cite{Davoudiasl:2011fj}.

%%%%%%%%%%%%%%%%%%%%%%%%%%%%%%%%%%%%%%%%%%%%%%%%
\begin{figure}[ttt]
\begin{center}
\if\withFigures1
\includegraphics[scale=.9]{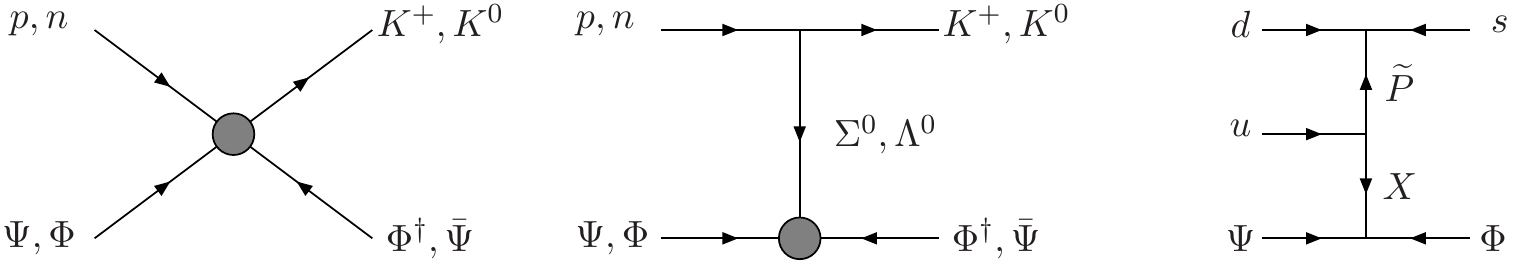}
\fi
\end{center}
\caption{IND processes at leading order in chiral effective theory (left, center). Gray dot shows effective $B$ transfer operator, generated by $\widetilde P$, $X$ exchange in our model (right).}
\label{fig:INDfigs}
\end{figure}
%%%%%%%%%%%%%%%%%%%%%%%%%%%%%%%%%%%%%%%%%%%%%%%%%%%%

Next, we compute the IND rates within our supersymmetric model, starting from the baryon transfer superpotential in Eq.~\eqref{UVmodel}.  The vector-like squarks $\widetilde P, \widetilde{P}^c$ mix through the SUSY-breaking term $\mathscr{L}_{\rm soft} \supset b_P \widetilde P \widetilde{P}^c$ to generate the mass eigenstates $\widetilde P_{1,2}$, with masses $m_{\widetilde P_{1,2}}$.  The leading contribution to IND arises from tree-level $\widetilde P_{1,2}$ exchange, shown in Fig.~\ref{fig:INDfigs}, giving (in two-component notation)
\beq
\mathscr{L}_{\rm eff} = \frac{1}{\Lambda_{\rm IND}^3}  \times \left\{ \begin{array}{ll} 
\epsilon_{\alpha\beta\gamma}(d_R^\alpha s_R^\beta)(u_R^\gamma \Psi_R)\Phi &  \quad \textrm{(case I)} \\ 
\epsilon_{\alpha\beta\gamma}(s_R^\alpha u_R^\beta)(d_R^\gamma \Psi_R)\Phi &  \quad \textrm{(case II)} \\
\epsilon_{\alpha\beta\gamma}(u_R^\alpha d_R^\beta)(s_R^\gamma \Psi_R)\Phi &  \quad \textrm{(case III)} 
\end{array} \right. , \;\; \frac{1}{\Lambda_{\rm IND}^3} \equiv \sum_{a=1,2} \frac{ 2 \bar\zeta_a^*  Z_{31} V^*_{11} b_P \blambda_a \blambda}{ m_{\widetilde P_1}^2 m_{\widetilde P_2}^2 m_{X_a}} \, . \label{INDquark}
\eeq
Here, we have neglected higher derivative terms, and $\Lambda_{\rm IND}$ characterizes the IND mass scale.  The different cases, corresponding to different baryon transfer interactions in Eq.~\eqref{UVmodel}, lead to different fermion contractions.  

The effective IND rate for nucleon $N = p,n$ is
\begin{align}
\Gamma(N \to K) = n_\Psi (\sigma v)_{\rm IND}^{N \Psi \to K \Phi^\dagger} + n_\Phi (\sigma v)_{\rm IND}^{N \Phi \to K \bar\Psi} 
\end{align}
where $n_{\Psi,\Phi}$ are the local DM number densities and $(\sigma v)_{\rm IND}$ is the IND cross section.  The IND lifetime can be expressed as
\beq
\tau(N \to K) = \frac{1}{\Gamma(N \to K)} =  \frac{(1+r)(\Omega_{\rm DM}/\Omega_{\rm b}) m_p }{2 \rho_{\rm DM} \big[  r (\sigma v)_{\rm IND}^{N \Psi \to K \Phi^\dagger} + (\sigma v)_{\rm IND}^{N \Phi \to K \bar\Psi} \big]}  
\eeq
with local DM mass density $\rho_{\rm DM} = m_\Psi n_\Psi + m_\Phi n_\Phi$, and assuming the local ratio $r \equiv n_\Psi/n_\Phi$ is the same as over cosmological scales.  The IND cross section is estimated as
\beq
(\sigma v)_{\rm IND} \approx \frac{m_{\rm QCD}^4}{16\pi \Lambda_{\rm IND}^6} \approx 10^{-39} \, {\rm cm}^3/{\rm s} \times \left( \frac{\Lambda_{\rm IND}}{1 \, \rm TeV} \right)^{-6} \; , \label{INDestimate}
\eeq
with QCD scale $m_{\rm QCD} \approx 1$ GeV.\footnote{The cross section $(\sigma v)_{\rm IND}$ also depends on the DM masses $m_{\Psi,\Phi}$, which for the purposes of dimensional analysis are comparable to $m_{\rm QCD}$.}  For $r \sim \mathcal{O}(1)$, the IND lifetime is
\beq
\tau(N \to K) \approx 10^{32} \, \textrm{yrs} \times \left(\frac{ (\sigma v)_{\rm IND} }{10^{-39} \, {\rm cm}/{\rm s} }\right)^{-1} \left(\frac{\rho_{\rm DM}}{0.3 \, {\rm GeV/cm}^{3} } \right)^{-1} \; ,
\eeq
which is exactly in the potential discovery range of existing nucleon decay searches, provided the baryon transfer scale $\Lambda_{\rm IND}$ is set by the weak scale.

%%%%%%%%%%%%%%%%%%%%%%%%%%%%%%%%%%%%%%%%%%%%%%%%
\begin{figure}[ttt]
\begin{center}
\if\withFigures1
\includegraphics[scale=.9]{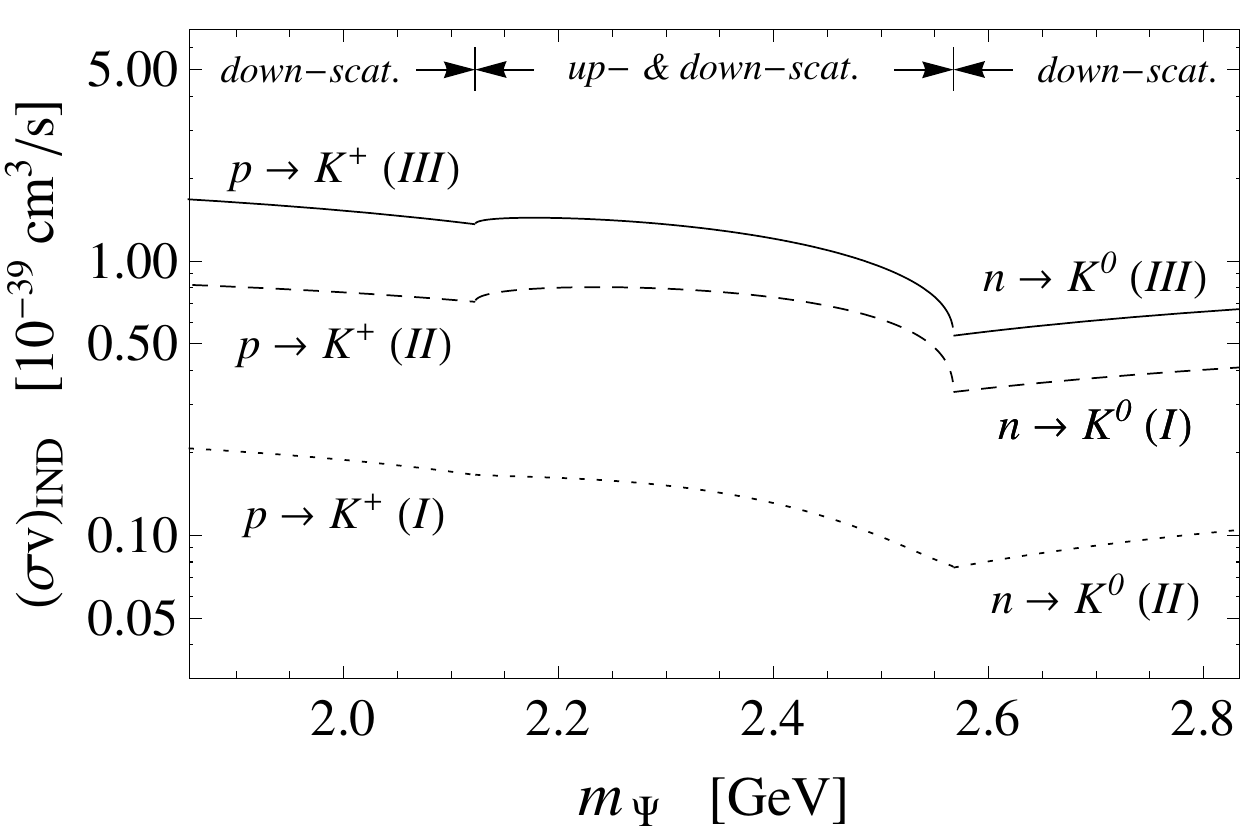}
\fi
\end{center}
\caption{Total IND cross section $(\sigma v)_{\rm IND} = (\sigma v)_{\rm IND}^{N \Psi \to K \Phi^*} + (\sigma v)_{\rm IND}^{N \Phi \to K \bar\Psi}$ over allowed range for $m_\Psi$, with $m_\Phi = (\Omega_{\rm DM}/\Omega_{\rm b}) m_p - m_\Psi \approx 5 m_p - m_\Psi$.  The effective baryon transfer mass scale is $\Lambda_{\rm IND} = 1$ TeV.  Cases I, II, III correspond to different baryon transfer models considered in  Eqs.~(\ref{UVmodel},\ref{INDquark}).}
\label{fig:INDrate1}
\end{figure}
%%%%%%%%%%%%%%%%%%%%%%%%%%%%%%%%%%%%%%%%%%%%%%%%%%%%

More quantitatively, we compute $(\sigma v)_{\rm IND}$ using chiral perturbation theory, which provides an effective theory of baryons and mesons (and DM) from the underlying quark-level interaction in Eq.~\eqref{INDquark}, following the same methods applied to standard nucleon decay~\cite{Claudson:1981gh} and with additional input from lattice calculations of hadronic matrix elements~\cite{Aoki:2008ku}. We refer the reader to Ref.~\cite{Davoudiasl:2011fj} for further details. Fig.~\ref{fig:INDrate1} shows numerical results for the total cross section $(\sigma v)_{\rm IND}$, over the allowed mass range $m_\Psi$ for $r=1$, for the three types of interactions in Eq.~\eqref{INDquark}.\footnote{The total rate $n \to K^0$ includes both $K^0_S$ and $K^0_L$ final states, and the individual channels $n \to K^0_S$ and $n \to K^0_L$ are (approximately) half the total rate.}  This calculation agrees well with our previous estimate in Eq.~\eqref{INDestimate}. However, since the typical IND momentum  is comparable to the chiral symmetry breaking scale $\approx 1 \, \textrm{GeV}$ (i.e., where the effective theory breaks down), we regard these results as approximate at best.  The different rates for different cases (for fixed $\Lambda_{\rm IND}$) satisfy
\beq
(\sigma v)_{\rm IND,III}^{p \to K^+} = (\sigma v)_{\rm IND,III}^{n \to K^0} \, , \quad (\sigma v)_{\rm IND,I}^{p \to K^+} = (\sigma v)_{\rm IND,II}^{n \to K^0} \, , \quad (\sigma v)_{\rm IND,II}^{p \to K^+} = (\sigma v)_{\rm IND,I}^{n \to K^0} 
\eeq
as a consequence of strong isospin symmetry~\cite{Davoudiasl:2011fj}.  The kinks correspond to up-scatting kinematic thresholds; to the left and right, only down-scatting is allowed, while in the center both up- and down-scattering occur.

%%%%%%%%%%%%%%%%%%%%%%%%%%%%%%%%%%%%%%%%%%%%%%%%
\begin{figure}[ttt]
\begin{center}
\if\withFigures1
\includegraphics[scale=.8]{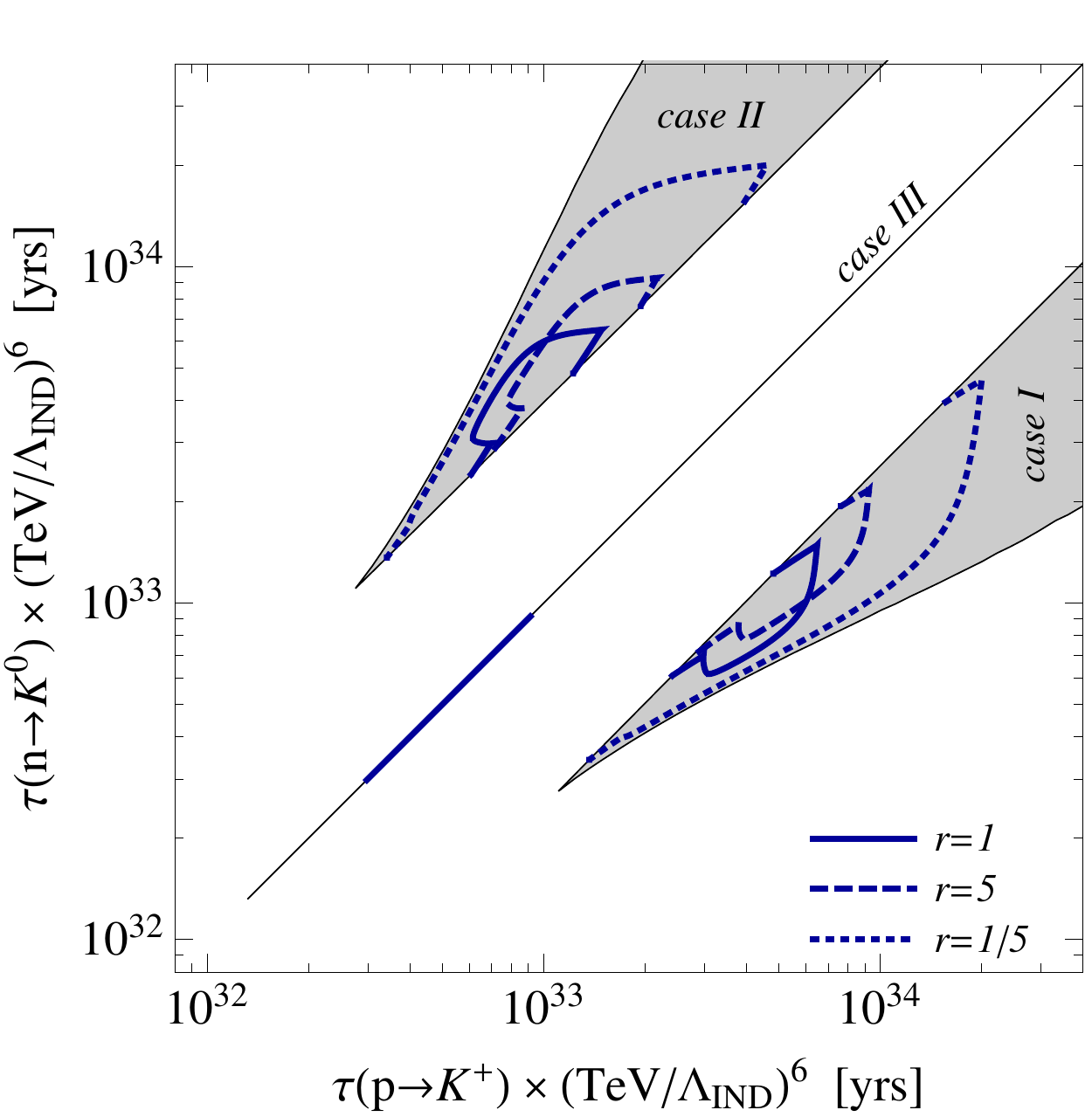}
\fi
\end{center}
\caption{(color online) Proton and neutron lifetimes for different baryon tranfer models (cases I, II, III) considered in  Eqs.~\eqref{UVmodel} and \eqref{INDquark}. Black line/gray regions show lifetime range for any $r$, while blue curves correspond to particular $r$ values.}
\label{fig:INDrate2}
\end{figure}
%%%%%%%%%%%%%%%%%%%%%%%%%%%%%%%%%%%%%%%%%%%%%%%%%%%%

More generally, we show in Fig.~\ref{fig:INDrate2} the allowed range for $p \to K^+$ and $n \to K^0$ IND lifetimes.  We consider masses $m_{\Psi,\Phi}$ consistent with $\Omega_{\rm DM}/\Omega_{\rm b} \approx 5$, for arbitrary $r$ in the range $0 < r < \infty$, and we take $\Lambda_{\rm IND} = 1$ TeV.   For case I (II), the allowed region is shown in gray, with a smaller (larger) lifetime for $n \to K^0$ than $p \to K^+$.  For case III, shown by the black line, the $p$ and $n$ IND lifetimes are equal, modulo SUSY radiative corrections and isospin-breaking that we neglect. Specific values for $r$ are shown by blue curves.  The solid blue curves show the IND lifetimes for $r=1$, corresponding to the calculation in Fig.~\ref{fig:INDrate1}.  For $r \ne 1$, the IND rate can be enhanced if the heavier state is overpopulated ({\it e.g.}, $r > 1$ for $m_\Psi > m_\Phi$ or $r < 1$ for $m_\Psi < m_\Phi$); on the other hand, the IND rate can be highly suppressed if the heavier state is depleted and up-scattering is kinematically forbidden.  The dashed (dotted) blue curves show the IND lifetimes for $r=5$ ($r=1/5$).  

We make a number of comments:
\begin{itemize}
\item The IND lifetimes scale as $\tau(N \to K) \propto \Lambda_{\rm IND}^6$.  Taking $\Lambda_{\rm IND}$ in the range $500 \, \textrm{GeV} - 5 \, {\rm TeV}$ corresponds to lifetimes of $10^{30} - 10^{36}$ years.   Lifetimes that can be probed in nucleon decay searches correspond to energy scales accessible in colliders (see below).
\item Both channels $p \to K^+$ and $n \to K^0$ provide complementary information, and which one dominates depends on the underlying heavy states mediating baryon transfer.
\item The largest IND rates in Fig.~\ref{fig:INDrate2} correspond to $\Phi$-dominated DM ($r \ll 1$) with $m_\Phi > m_\Psi$, and IND is dominated by $\Phi N \to \bar\Psi K$ down-scattering.
\end{itemize}
Lastly, we note that while the observation of IND would be a smoking gun signal for hylogenesis, a nonobservation does not rule out hylogenesis as a baryogenesis mechanism.  The IND rate can be suppressed if (i) the effective scale $\Lambda_{\rm IND}$ lies beyond the TeV-scale (due to small couplings or large mass parameters), (ii) baryon transfer in the early Universe involves heavier quark flavors, and/or (iii) the heavier DM state is depleted while up-scattering IND is kinematically blocked.

\subsection{Precision Probes}
 
At energies well below the weak scale, the light states in the hidden 
sector interact with the SM primarily through the gauge kinetic mixing 
interaction.  The most important effect of this mixing is an induced
coupling of the $Z'$ vector boson to the SM fermions $f$ given by
\beq
-\mathscr{L} \supset -\kappa c_WQ_f^{em}\bar{f}\gamma^{\mu}Z'_{\mu}f \ ,
\label{photonmix}
\eeq  
where $Q_f^{em}$ is the electric charge of the fermion and $c_W$ is the
cosine of the weak mixing angle.  Direct searches for a light $Z'$ limit 
$\kappa c_W \lesssim 10^{-3}$ for $m_{Z'} \lesssim 1\gev$~\cite{
Pospelov:2008zw,Bjorken:2009mm,Reece:2009un} with significant 
improvements expected in the coming 
years~\cite{Merkel:2011ze,Abrahamyan:2011gv,Andreas:2010tp}.  

  The dark matter states in our scenario consist of a Dirac fermion or 
a complex scalar with a direct coupling to the $Z'$ vector.  With the
mixing interaction connecting the $Z'$ to SM fermions, this state will
efficiently mediate spin-independent elastic scattering of the DM states 
off nuclei.  We estimate the cross section per target 
nucleon to be~\cite{Davoudiasl:2010am}
\bea
\sigma_0^{SI} = (5\times 10^{-39} \, {\rm cm}^2)\lrf{2Z}{A}^2\lrf{\mu_{n}}{\gev}^2
\lrf{e'}{0.05}^2
\lrf{\kappa}{10^{-4}}^2
\lrf{0.3\, \gev}{m_{Z'}}^4,
\eea
where $\mu_n$ is the DM-nucleon reduced mass.
While this cross section is quite large, the masses of the DM particles
in this scenario lie below the region of sensitivity of most current 
direct detection DM searches, including the specific low-threshold analyses by
COGENT~\cite{Aalseth:2010vx}, CDMS~\cite{Akerib:2010pv}, 
XENON10~\cite{Angle:2011th}, and XENON100~\cite{Aprile:2011hi}. 
For a DM mass below $3\,\gev$, this cross section lies slightly below the 
current best limit from CRESST~\cite{Angloher:2002in}. 
Proposed low-threshold searches for DM scattering with nuclei or electrons
are expected to improve these limits~\cite{Essig:2011nj}.

\subsection{High-Energy Colliders}

The new heavy states required for hylogenesis couple directly to the 
SM and can potentially be probed in high-energy colliders such as the
Tevatron and the LHC.  In particular, the effective interactions
induced by the vector-like quark multiplets $(P,P^c)$ can generate monojets
and modify the kinematic distributions of dijets.  We discuss here the
approximate limits that existing collider data places on the masses
of these multiplets, although we defer a detailed analysis to the future.  

Monojet signals arising from the effective four-fermion interaction
$(X u_R^c d_R^cs_R^c) /M^2$ present in the minimal hylogenesis model were considered previously in Ref.~\cite{Davoudiasl:2011fj}.  More recent searches 
for monojets by the ATLAS~\cite{atlas-monoj} and CMS~\cite{Aaltonen:2012jb} 
collaborations limit the corresponding mass scale $M$ to lie above $0.5-3$ TeV.  
Note, however, that in our supersymmetric formulation the corresponding 
four-fermion operator is only generated once supersymmetry-breaking effects are included.  This weakens the correlation between the monojet signal and the operator responsible for hylogenesis, although the limit does typically force the $P^{(c)}$ multiplets to be at least as heavy as a few hundred GeV.  On the other hand, this operator is directly related to the IND interaction.  An alternative signal that can arise directly from the superpotential interaction is a ``monosquark'' $\tilde{q}^*\widetilde{X}$ final state, with the squark decaying to 
a jet and missing energy.  In both cases, collider limits may be weakened 
through cascade decays in the hidden sector, which could produce additional hidden photons or Higgs bosons that decay to SM states.

  A second way to probe our supersymmetric UV completion of hylogenesis
is through the kinematic distributions of dijets, which can be modified 
by the direct production of the triplet $P$ scalars (which are $R$-even).  
On-shell production of scalar $\tilde{P}$ states via the interactions 
of Eq.~\eqref{UVmodel} can produce a dijet resonance.  For heavier masses, 
the primary effect is described by the non-minimal K\"{a}hler potential 
operator of Eq.~\eqref{newkahler}, which gives rise to a four-quark contact 
operator.  Studies of dijet distributions by ATLAS~\cite{atlas-qcontact} and 
CMS~\cite{Chatrchyan:2011ns,Chatrchyan:2012bf} put limits on the masses
of the $P^{(c)}$ scalars of 1-10 TeV, although the specific limits depend
on the flavour structure of the quark coupling in Eq.~\eqref{UVmodel} 
present in the underlying theory.

\section{Conclusions\label{sec:conc}}

Through the mechanism of hylogenesis, the cosmological densities of visible and dark matter may share a unified origin.  Out-of-equilibrium decays during a low-temperature reheating epoch generate the visible baryon asymmetry, and an equal antibaryon asymmetry among GeV-scale hidden sector states.  The hidden antibaryons are weakly coupled to the SM and are the dark matter in the Universe.  

%We have shown that SUSY provides a natural framework for hylogenesis. 
We have embedded hylogenesis in a supersymmetric framework.  
By virtue of its weak couplings to the SM, SUSY-breaking is sequestered from the hidden sector, thereby stabilizing its GeV mass scale. The DM consists of two states, a quasi-degenerate scalar-fermion pair of superpartners.  We studied in detail one particular realization of supersymmetric hylogenesis, considering several aspects:
\begin{itemize}
\item We constructed a minimal supersymmetric model for hylogenesis.  Hidden sector baryons are chiral superfields $X$ and $Y$, with $B=1$ and $-1/2$, respectively. The lightest $Y$ states are DM, while $X$ decays in the early Universe generate the $B$ asymmetries.  
\item In addition, we introduced a vector-like $SU(3)_C$ triplet to mediate $B$ transfer between visible and hidden sectors, and a hidden $Z^\prime$ gauge boson (with kinetic mixing) to deplete efficiently the symmetric DM densities.
\item We showed that hylogenesis can successfully generate the observed $B$ asymmetry during reheating.  We computed the $\cp$ asymmetry from $X$ decay and solved the coupled Boltzmann equations describing the cosmological dynamics of hylogenesis.  
\item We studied how SUSY breaking is communicated between the visible and hidden sectors through RG effects.  We also examined predictions within an AMSB framework. While anomaly mediation explains the late-time reheating epoch from moduli decay, we have not explicitly addressed the issues of tachyonic slepton masses 
in the visible sector and the origin of SUSY mass terms in the hidden sector.
\item Antibaryonic DM annihilates visible nucleons, causing induced nucleon decay to kaon final states, with effective nucleon lifetime in the range $10^{30} - 10^{36}$ years.  DM can be discovered in current nucleon decay searches, and this signal remains unexplored.
\item Collider searches for monojets and dijet resonances provide the strongest direct constraints on our model, and these signals are correlated with IND.  Lifetimes of $10^{30} - 10^{36}$ years correspond to energy scales $\Lambda_{\rm IND} \sim 0.5 - 5$ TeV that can be probed at the LHC.
\item DM direct detection experiments and precision searches for hidden photons constrain the $Z'$ kinetic mixing, although our model remains consistent with current bounds.
\end{itemize}
We emphasize that our specific model was constructed to illustrate general features of hylogenesis, and certainly there are many other model-building possibilities along these lines.  Nevertheless, it is clear that supersymmetric hylogenesis provides a technically natural and viable scenario for the genesis of matter, explaining the cosmic coincidence between the dark and visible matter densities and predicting new experimental signatures to be explored in colliders and nucleon decay searches.

\section*{ Acknowledgements}

We thank K.~Zurek for helpful discussions.
DM and  KS would like to thank Perimeter Institute for 
Theoretical Physics for their hospitality.
The work of NB, DM, and KS is supported by the National Science
and Engineering Research Council of Canada~(NSERC).
The work of ST was supported in part by the United States Deparment of Energy
grant \#DE-FG02-95ER40899 and by NSERC.

\appendix 

\section{Alternate Higgs portal model}
\label{sec:hportal}

We present here an alternate model for hylogenesis.  In this model, hylogenesis cosmology is the same as described in Sec.~\ref{sec:baryo}: a nonthermal abundance of $X_1$ states arises during a low temperature reheating epoch and decays out of equilibrium to produce equal-and-opposite baryon asymmetries in the visible and hidden sectors.  The main difference, compared to the model in Sec.~\ref{sec:susy}, is that the symmetric DM particle-antiparticle density annihilates into light (GeV-scale) scalars, which in turn decay to SM fermions through mixing with the MSSM Higgs pseudoscalar $A^0$.  We do not introduce a hidden sector $U(1)^\prime$ gauge symmetry, and this allows for a reduced field content.

We suppose the hidden sector consists of (i) three vector-like chiral supermultiplets $X_{1,2}$ and $Y$, carrying $B=+1$ and $B=-1/2$, with charge-conjugate partners $X_{1,2}^c$ and $Y^c$, respectively; and (ii) a single $B=0$ chiral supermultiplet $N$.
%\footnote{This model has three fewer chiral supermultiplets compared to the one given in Sec.~\ref{sec:susy}, which has ten chiral supermultiplets in total ($X_{1,2},X^c_{1,2}, Y_{1,2}, Y^c_{1,2}, H, H^c$).}  
The $X_{1,2}$ states are coupled to SM quarks via Eq.~\eqref{trans}, and are responsible for generating $\cp$ asymmetries and $B$ transfer between visible and hidden sectors, as in Sec.~\ref{sec:susy}.  The hidden sector superpotential is  
\begin{align}
W_{\rm HS} = \zeta_a X_a Y^2+ \varepsilon \, N H_u H_d + \xi N Y Y^c + \frac{k}{3} \, N^3 
+ \mu_{X_a} X_a X^c_a \; , \label{altWHS}
\end{align}
with couplings $\zeta_{1,2}$,  $\varepsilon$, $\xi$, $k$, and where $H_{u,d}$ are the usual MSSM Higgs supermultiplets.  Additional terms are forbidden by introducing a global $\mathbb{Z}_3$ symmetry under which $N$ and $Y^c$ carry opposite charge, while other fields are neutral.  Since $N H_u H_d$ breaks this symmetry explicitly, we take $\varepsilon \ll 1$.
The soft SUSY-breaking terms are
\begin{align}
- \mathscr{L}_{\textrm{soft}} \supset  m_N^2 |N|^2 + m_{\widetilde Y}^2 |\widetilde Y|^2 + m_{\widetilde Y^c}^2 |\widetilde Y^c|^2  +  \big( A_k N^3 + A_\varepsilon N H_u H_d + A_\xi N \widetilde{Y} \widetilde{Y}^c + \textrm{h.c.} \big) \; . \label{altsoft}
\end{align}
We assume all couplings are real, and take $m_N^2 < 0$ such that the scalar component of $N$ acquires a vev $\langle N\rangle \equiv \nu$, which generates mass terms for the DM states.

The physical spectrum of the hidden sector is given as follows.  From the $(Y,Y^c)$ supermultiplet pair, we have a Dirac fermion 
$\Psi \equiv (Y,Y^{c\dagger})$, with mass $m_\Psi = |\xi  \nu |$, and two scalar states $\widetilde Y, \widetilde Y^c$, described 
by the mass matrix
 \beq
\mathscr{L}_{\textrm{mass}} = - \left(\widetilde Y^\dagger, \, \widetilde Y^c\right) \left( \begin{array}{cc} m_{\widetilde Y}^2 + m_\Psi^2 &
\xi k \nu^2 + A_\xi \nu \\ \xi k \nu^2 + A_\xi \nu & m_{\widetilde Y^c}^2 + m_\Psi^2 \end{array} \right)\left( \begin{array}{c} \widetilde Y \\ \widetilde Y^{c\dagger} \end{array} \right) \; .
\eeq
The mass eigenstates are given by $\Phi_{1,2}$, with masses
\beq
m_{\Phi_{1,2}}^2 = \frac{1}{2} \left( 2\, m_\Psi^2 + m_{\widetilde Y}^2 + m_{\widetilde Y^c}^2 \pm \sqrt{ (m_{\widetilde Y}^2 - m_{\widetilde Y^c}^2)^2 + 4 \,(\xi k \nu^2 + A_\xi \nu)^2 } \right) \; ,
\eeq
defined such that $m_{\Phi_1}^2 < m_{\Phi_2}^2$.  DM consists of the states $\Psi$ and $\Phi_1$.  For suppressed soft masses ($m_{\widetilde Y,\widetilde Y^c}^2 \ll m_\Psi^2$), mixing is maximal (the relevant mixing angle is $\theta \approx \pi/4$) and $m_{\Phi_1} < m_\Psi < m_{\Phi_2}$.  

The spectrum of $N$ states consists of a real scalar $s$, pseudoscalar $a$, and Majorana fermion $\widetilde N$.
In the limit $\varepsilon \ll 1$, the leading $\mathcal{O}(\varepsilon)$ effects will be mixing between these states and the MSSM Higgs bosons and Higgsinos.   Here, we set $\varepsilon = 0$, obtaining
\beq \label{massrel}
m_s^2 = 4 k^2 \nu^2 + A_k \nu \, , \quad m_a^2 = - 3 A_k \nu \, , \quad m_{\widetilde N}^2 = 4 k^2 \nu^2 \; ,
\eeq
where the soft mass $m_N^2$ has been traded for $\nu$ using the minimization condition.  
Vacuum stabilility requires $m_{s,a}^2 > 0$.  These relations can be combined: $m_{\widetilde N}^2 = m_s^2 + \frac{1}{3} m_a^2$.  

A light pseudoscalar ($m_a < m_{\Psi}, m_{\Phi_1}$) can provide an efficient channel for DM annihilation $\Psi \bar\Psi \to aa$ and $\Phi_1 \Phi^\dagger_1 \to aa$, with cross sections $\langle \sigma v \rangle \gtrsim 3 \times 10^{-26} \, \textrm{cm}^3/\textrm{s}$ required to deplete the symmetric DM densities.  For $\Phi_1$ annihilation, the interaction $\mathscr{L} \supset - \frac{\xi^2}{2} \, a^2 \, |\Phi_1|^2$ gives
\beq
\langle \sigma v \rangle_{\Phi_1 \Phi^\dagger_1 \to a a} = \frac{\xi^4}{256\pi \, m_{\Phi_1}^2} \, \sqrt{1 - m_a^2/m_{\Phi_1}^2 } \approx 4 \times 10^{-25} \, \textrm{cm}^3/\textrm{s} \times \left( \frac{\xi}{0.1} \right)^4 \left( \frac{m_{\Phi_1}}{2 \; \textrm{GeV}} \right)^{-2} \; .
\eeq
For $\Psi$ annihilation, $t$-channel $\Psi \bar\Psi \to a a$ is $p$-wave suppressed; however, if $s$ is not too heavy ($m_s \lesssim 20$ GeV), it can mediate efficient $s$-channel annihilation:
\beq
\langle \sigma v \rangle_{\Psi \bar\Psi \to a a} = \frac{\xi^2 (A_k - 2 k^2 \nu)^2}{256\pi\, (m_s^2 - 4 m_\Psi^2)^2} \sqrt{1 - {m_{a}^2}/{m_\Psi^2} } \; \approx \;
\; \frac{\xi^4 m_{\widetilde S}^4}{1024 \pi \, m_\Psi^2 (m_s^2 - 4 m_\Psi^2)^2} \; ,
\eeq
where the second step follows by using the mass relations given in Eq.~\eqref{massrel} and assuming $m_a \approx 0$.  For $m_\Psi = 3$ GeV, $m_s = m_{\widetilde S} = 10$ GeV, $\xi = 0.1$, we have $\langle \sigma v\rangle \approx 4 \times 10^{-25} \, \textrm{cm}^3/\textrm{s}$.  
In both cases, as long as $\xi \gtrsim 0.1$, the cross sections can easily be large enough to deplete the symmetric densities.

Depletion of the heavier DM state, {\it e.g.} $\Psi \bar \Psi \to \Phi_1 \Phi^\dagger_1$ for $m_{\Phi_1} < m_\Psi$, may occur through $t$-channel $\widetilde N$ exchange.  We require $m_{\widetilde N} > m_{\Phi_1} + m_{\Psi} \sim 5$ GeV to allow for decays $\widetilde N \to \Phi_1 \bar\Psi$ (otherwise $\widetilde N$ would be an additional stable DM component); however $\widetilde N$ cannot be too heavy since $m_{\widetilde N} \approx m_s \lesssim 20$ GeV.  For $m_{\widetilde S} \gg m_\Psi - m_{\Phi_1}$, and neglecting terms that are $p$-wave suppressed, the leading contribution to this cross section is
\beq
\langle \sigma v \rangle = \frac{\xi^4 \cos^2 2\theta}{64 \pi \, m_{\widetilde S}^2} \, \sqrt{1 - m_{\Phi_1}^2/m_\Psi^2 } \; ,
\eeq
where $\theta$ is the $\widetilde Y$ mixing angle.  For $m_\Psi = 3$ GeV, $m_{\Phi_1} = 2$ GeV, $m_{\widetilde S} = 10$ GeV, $\xi = 0.1$, we have
\beq
\langle \sigma v \rangle \approx 4 \times 10^{-26} \, \textrm{cm}^3/\textrm{s} \times \cos^2 2\theta \; .
\eeq
The cross section for depletion of the heavier DM component is suppressed for maximal mixing ($\theta \approx \pi/4$), corresponding to the case when $m_{\widetilde Y}^2, m_{\widetilde Y^c}^2 \ll m_{\Psi}^2$.

After DM annihilation, the pseudoscalar can decay to SM fermions via mixing with the MSSM pseudoscalar boson $A^0$ through the $\varepsilon N H_u H_d$ term.  The decay rate is
\beq
\Gamma(a \to f \bar{f}) = \frac{ N_c \sin^2 \alpha \, m_f^2 m_a}{ 16\pi \, v^2} \, \sqrt{ 1- 4m_f^2/m_a^2 } \times \left\{ \begin{array}{cc} \cot^2 \beta & f = u \\ \tan^2 \beta & f = d,\ell \end{array} \right.
\eeq
where the $A^0$-$a$ mixing angle is $\alpha \approx - \varepsilon m_{\widetilde S} v/(2 \, m_{A^0}^2)$.  To avoid BBN constraints, we want $\Gamma^{-1} \lesssim 1 \, \textrm{s} \approx 6\times 10^{-25} \, \textrm{GeV}^{-1}$.  For $a \to \mu \bar\mu$, we have
\beq
\Gamma(a \to \mu \bar\mu) \approx 7 \times 10^{-9} \, \textrm{GeV} \times \tan^2\beta \sin^2 \alpha \, \left(\frac{m_a}{1 \, \textrm{GeV} }\right) \; .
\eeq
The decay rate for $a \to s \bar{s}$ is faster by a factor $\mathcal{O}(3)$ if kinematically available.

%%%%%%%%%%%%%%%%%%%%%%%%%%%%%%%%%%%%%%%%%%%%%%%%%%%%%%%%%%%%%%%%%%%%%%

%\newpage

\end{document}